\documentclass[11pt, a4paper]{article}

\usepackage[T1]{fontenc}
\usepackage{lmodern}
\usepackage[margin=1in]{geometry}

\usepackage{hyperref}
\usepackage{xspace}
\usepackage{color,soul}
\usepackage{enumerate}

\usepackage[round]{natbib}

\usepackage{amsmath, amssymb, amsthm}

\usepackage{graphicx}
\usepackage{tikz}
\usetikzlibrary{calc}
\usepackage{subcaption}
\usepackage{booktabs}
\usepackage{setspace}
\usepackage{comment}
\usepackage{fancyhdr}
\usepackage{algorithmic}
\usepackage{textcomp}
\usepackage{upgreek}
\usepackage{bm}
\usepackage[linesnumbered, ruled]{algorithm2e}

\theoremstyle{plain}
\newtheorem{proposition}{Proposition}
\newtheorem{theorem}{Theorem}
\newtheorem{lemma}{Lemma}
\newtheorem{corollary}{Corollary}
\newtheorem{assumption}{Assumption}

\theoremstyle{definition}

\newtheorem{definition}{Definition}

\usepackage{fancyhdr}

\DeclareMathOperator*{\argmax}{arg\,max}

\title{Effect of a Manager in Relational Contracts with Multiple Workers\\

 \normalsize{Manager in Relational Contracts}}
\author{Beomjun Kim \thanks{B. Kim is with the Department of Mechanical Engineering, Massachusetts Institute of Technology, Cambridge, MA, USA} \thanks{I acknowledge the supervision of Professor Yves Gueron in Seoul National University and many helpful advices given by Professor Ilwoo Hwang in Seoul National University,} }
\date{April 2025}
\begin{document}
\begin{titlepage}
\maketitle

\begin{abstract}
This paper considers the optimal management structure about hiring a manager and providing the manager with a separate salary and bonus using a relational contract among an owner, a manager, and workers, assuming that the manager can observe individual worker performances while the owner can observe only overall team performance. I derive optimal contracts for the two cases in which the manager's salary and bonus are integrated into total team bonus or provided separately. I compare situations of having the manager distribute bonuses based on individual worker performance to the situation of equal bonus distribution based on overall team performance without a manager. Only a contract with a manager who receives a separate bonus is feasible for low discount factor. Making the manager to distribute the salary and bonus including himself is best with intermediate discount factor. Providing an equal bonus without a manager is optimal with high discount factor. 

\end{abstract}

{\bf JEL Classification: D86, D23}

{\bf Keywords:} Relational contract, Manager, Monitoring, Optimal management structure
\end{titlepage}

\begin{abstract}
This paper considers the optimal management structure about hiring a manager and providing the manager with a separate salary and bonus using a relational contract among an owner, a manager, and workers, assuming that the manager can observe individual worker performances while the owner can observe only overall team performance. I derive optimal contracts for the two cases in which the manager's salary and bonus are integrated into total team bonus or provided separately. I compare situations of having the manager distribute bonuses based on individual worker performance to the situation of equal bonus distribution based on overall team performance without a manager. Only a contract with a manager who receives a separate bonus is feasible for low discount factor. Making the manager to distribute the salary and bonus including himself is best with intermediate discount factor. Providing an equal bonus without a manager is optimal with high discount factor. 
\end{abstract}
\section{Introduction}
Large enterprises exist everywhere. In the US, large enterprises are defined as those with 1,000 or more workers. According to the \cite{uscensus}, large enterprises employed more than half of all workers in the United States in 2012. The extant literature on relational
contract assumes contracts between a principal (owner) and agents. However, owners of large enterprises typically cannot observe all aspects of an agent's performance, which prevents them from efficiently contracting with agents. Conversely, when they hire a manager, as the manager is responsible for only a few assigned workers (typically team members), the manager can check individual performances better than the owner. Team bonuses can be distributed by team managers, divided equally or based on seniority among team members. Additionally, the manager's bonus can be integrated into the team bonus or given separately by the owner. In this paper, the research question is thus as follows: when will the owner prefer each contract mechanism among these management and bonus structures?

In this paper, I propose relational contract models of firm incentive problems by extending the well-known principal-agent multilateral contract model of \cite{10.1162/003355302760193968}. In reality, managers in firms generally work with their team members or have daily meetings with them. Moreover, in most cases, managers are the first approvers of their team members. Thus, I intuitively assume that managers can observe the individual performance of their team members, as in \cite{10.1162/003355302760193968} and Section 3 of \cite{kvaloy2019relational}. Conversely, an owner usually does not interact with all of the firm's workers. Therefore, in most cases, I can assume that the owner cannot observe the performance of the firm's workers. For simplicity, I consider only one manager and an owner who can observe the overall performance of a team.

In most firms, managers evaluate their team member's performance, and the evaluation is used as the main source of data for rewarding bonuses. Even though bonuses are rewarded in the name of the owner, the de facto decision maker in terms of the bonus is the manager because managers can manipulate the evaluation of their team members' performance and reward bonuses as they wish. Conversely, the total bonus for team members is generally limited by the owner based on overall team performance. Therefore, I assume that the owner assigns the total team bonus to the manager, and the manager distributes the bonus at his or her discretion. At this point, it can be a major difference in the model whether the manager's salary and bonus are included in the team's salary and bonus. If the manager's salary and bonus are included in the team's salary and bonus, respectively, the manager can take the residual bonus, and the contract is similar to the subcontract system. If the manager's salary and bonus are given separately by the owner, the manager's discretion is limited to the distribution of the salary and the bonus. However, the manager can collude with a low-performance worker, and together, they can pocket the bonus intended for the high-performance worker. I include a parameter that represents the ease of such collusion. I solve these two cases separately and compare the results.

I derive the optimal bonus scheme of the models via backward induction. I first solve the contract between the manager and the workers regarding the distribution of bonuses (denoted as the \textit{inner contract}) with an approach similar to \cite{10.1162/003355302760193968,RePEc:hhs:nhhfms:2016_023,kvaloy2019relational}. I construct the optimization problem of the manager under participation and incentive constraints of workers and solve it using Lagrangian duality. Then, I find the optimal solution for whole contracts. In both contracts, the bonus scheme for the workers is the tournament mechanism with a threshold, that is, the best-performing worker receive the whole bouns provided that her performance is higher than the fixed threshold. In the separate bonus case, the threshold is smaller than the optimal threshold for effort induction to ensure workers' participation under a low salary. When the manager's salary and the bonus are integrated into the team bonus, the optimal effort consists of two parts due to the ability of the profit of the manager to enforce the participation without reneging. The first part is when the variance of the performance is sufficiently small and profit of the manager is sufficiently large for manager to continue the contract without refusing to give the bonus. In this part, the optimal soluation is the global optimal solution for the owner. The other part is when the profit of manager is not sufficeintly high and thus the total bouns the manager can give without incentive of reneging is limited. In this part, the enforcement constraint of the provision of the bonus by the manager and the owner binds. The optimal effort and the surlpus rapidly decreases for an increase in the variance in performance or the outside income of the manager. In the separate bonus case, since assigning the total bonus to the manager can be a commitment device, there is no constraint regarding the manager profit to give the bouns without reneging. However, because of the inefficient effort-inducing ability of the bonus mechanism, the optimal effort and surplus of the separate bonus constraint can be outperformed by the first part of the solution of integrated bonus case when there is no device to prevent the collusion of the manager with a worker.

I show that the result is simplified when the performance follows a normal distribution and the mean is the same as the effort. (This assumption is widely used in the literature \cite{LiangPierreJinghong2008OTSa,LiangPierreJinghong2014EPoP,kvaloy2019relational}) Then, I 

prove 
that the optimal ratio of the profit of the owner and the profit of the manager is constant under a normal distribution. I compare this model with the case in which bonuses are shared equally among all team members for the simplified case. The increase in the number of workers in a team makes the contracts with the manager more efficient; however, it makes the equal bonus contract more inefficient. This is because the effort-inducing ability of the equal bonus mechanism is inefficient and becomes more inefficient when the number of team members increases. I visualize the result under a simple cost function and environmental variables and analyze the effects of environmental variables.

Finally, I analyze which contract is the best choice for the owner among these three types of contracts (equal bonus, integrated bonus with manager, separate bonus for manager). The manager must be paid, but he or she can help to achieve greater effort if the first best effort cannot be achieved in the equally shared bonus case because of the small discount factor. Thus, the equally shared bonus scheme is better when the discount factor goes to $1$ or the variance goes to $0$. Conversely, when the variance is high or the discount factor is low, only hiring a manager and giving a separate salary and bonus to the manager is feasible. Hiring a manager with an integrated bonus is efficient for the middle-range value of variance (or discount factor) when the manager's outside income is low, and there is virtually no system to prevent collusion between a manager and a worker.

The remainder of this paper is organized as follows. In the remainder of this section, I present some related works. In Section \ref{sec_model}, I present the basic model of the problem. Section \ref{sec2}, I introduce and solve the model where the owner hires a manager and gives an integrated team bonus. In Section \ref{sep}, I introduce and solve the model where the owner hires a manager and gives a separate bonus to the manager. In Section \ref{sec4}, I simplify the model with a normal distribution function, visualize the result as a function of the parameters with an interpretation of the parameters, and present the optimal management structure for the owner.

\subsection{Related work}
Relational incentive contracts, which combine ideas developed in a one-shot contract model, were first introduced in the 1900s by \cite{HolmstromBengt1979MHaO,holmstrom1982moral,kandel1992peer}. They garnered renewed interest in the literature in the 2000s and since then have been extensively used to address moral hazard problems in firms \cite{10.2307/2696482,10.1162/003355302760193968,KvaloyOla2006TIiR,RayoLuis2007RIaM,RePEc:hhs:nhhfms:2016_023,kvaloy2019relational}. \cite{10.2307/2696482} use the relational contract concept to analyze why vertical integration is beneficial. \cite{10.1162/003355302760193968} compares bilateral and multilateral contracts in employment and shows that multilateral contracts are efficient but require a higher discount factor. \cite{KvaloyOla2006TIiR} use a simple model to analyze the conditions when joint, relative, and independent performance evaluation is best for rewarding bonuses. \cite{RayoLuis2007RIaM} discusses the effects of profit sharing in addition to a relational contract with a bonus. \cite{10.1257/aer.20131082,10.1257/mic.20170181,10.1257/aer.101.7.3349} considers workers or suppliers competing to be rewarded by a principal. Finally, \cite{RePEc:hhs:nhhfms:2016_023} and \cite{kvaloy2019relational} present how team size correlates with worker performance and affects the efficiency of a multilateral relational incentive contract. Other approaches, including empirical analysis to address the incentive problem in institutions, are well summarized in \cite{lazear2018compensation}.

Managerial effort and optimal team size are also discussed in the literature. \cite{NagarVenky2002DaIC} considers how much authority should be delegated to managers. \cite{LiangPierreJinghong2008OTSa} use a linear-exponential-normal (LEN) model to derive the optimal team size under environmental variables and analyze the effects of a manager's task specialization and monitoring efforts. \cite{LiangPierreJinghong2014EPoP} handle the optimal policy for shareholders considering the tradeoff between productive effort and performance-reporting effort. \cite{FuRichard2016PCIa} develop a model and analyze the effect of project risks, monitoring costs, and profitability on optimal team size. \cite{DurrOliver2020Iios} discuss the role managers play in reducing productivity risk and performance measurement noise in a case where all players are risk-averse. \cite{HofmannChristian2021AMaI} consider monitoring and providing incentives in the principal-manager-$n$ workers setting while also accounting for the effect of the delegation of contracting authority to the manager.

The proposed framework has several advantages over the research regarding the managerial effort and optimal team size mentioned above. First, the possibility of some players reneging, which is not considered in the single-period contract models used by \cite{LiangPierreJinghong2008OTSa,LiangPierreJinghong2014EPoP,FuRichard2016PCIa,DurrOliver2020Iios,HofmannChristian2021AMaI}, is addressed. Second, in the integrated bonus case, since the manager distributes the team bonus to maximize his or her profit, there is no possibility of false reporting in the proposed framework in contrast to the settings of \cite{LiangPierreJinghong2008OTSa} and \cite{DurrOliver2020Iios} in which the manager can falsely report on performance or other information. In the separate bonus case, the type of reneging is addressed as a part of the model (parameter $\phi$). Third, the proposed framework can cover a wider range of cost functions and the probability distribution of the performance than the simple linear contract model used by most of the literature. Finally, the optimal management structure (whether the owner hires a manager, whether the manager's bonus is integrated into the team bonus) can be derived as a function of parameters in addition to the analysis of the effects of each parameter (variance, the number of workers, etc.

Recently, similar to this paper, \cite{10.1093/jeea/jvac049} dealt with relational contracts with a manager. The paper analyzes how a relational contract managed by a manager differs from a relational contract between a principal and an agent. The paper focuses on the possibility of collusion and its drawbacks. The paper further addresses whether the owner (or principal) should or should not entrust the relationship to a manager. The main difference between \cite{10.1093/jeea/jvac049} and this paper is the type of contract, depth of mathematical analysis, and the comparison between the contracts. In contrast to \cite{10.1093/jeea/jvac049}, who analyze a bilateral contract, this paper deals with a multilateral contract in which the manager manages several team members. Moreover, this paper provides an analysis of the general function between effort and performance in the appendix. Finally, this paper provides a detailed comparison between the types of contracts the owner can choose and obtains conditions in which each type of contract is best for the owner.

\section{Model}\label{sec_model}
In this subsection, I introduce the problem that I address in this paper. I consider an infinitely-repeated stationary situation in which an owner employs $n$ symmetric workers. I assume $n$ as a given parameter since choosing or changing the management system occurs much frequently than the macroscopic change of the fiem size. Workers participate in the contract and choose their own hidden effort $e_i$ with cost $c(e_i)$ in each period if the contract is beneficial to them. Cost function $c(e_i)$ is strictly increasing, convex in effort, and $c(0)=c' (0)=0$. Managers always select workers' efforts to maximize their profit. The efforts lead to performance $\bold{x}=(x_1, \dots, x_n)$. (The relation between $x_i$ and $e_i$ is governed by a probability density function $f(x_i;e_i)$ and independent with each other). I allow the negative performance ($x_i<0$). In the main part, I only address the normal distribution centered on $e_i$. The general derivation of the optimal contract and analysis is provided in the appendix.) However, the owner cannot directly oversee the workers. Instead, the owner receives and freely observes revenue $y=\sum{x_i}$. The owner can employ a manager to oversee the workers and delegate the authority of distributing salary $\alpha_i$ and bonus $b_i$ to workers. The manager receives salary and bouns, both of them will be defined at the following sections \ref{sec2},\ref{sep}. Bonuses are constrainted to be positive and can depend on the performance. In sum, I consider an ongoing relational contract among an owner, possibly a manager, and $n$ symmetric workers. I assume risk neutrality for all parties and the stationarity of the contract over time because I consider it an infinitely repeated and optimal contract (see \cite{10.2307/3132119}). The common discount factor is denoted as $\delta$. I denote $\bar{u}$ and $\bar{u_0}[n]$ as the earnings of the workers and the manager when they are employed elsewhere. Note that a manager who can manage many workers may have a high outside income. Moreover, the cost of a system to manage many workers by one manager, as in the case of management information systems, also increases as the number of workers increases and can be considered as included in $\bar{u_0}[n]$ since the fixed cost can be transferred between the owner and the manager in the form of the manager's salary. Thus, $\bar{u_0}[n]$ will be an increasing function of $n$. I assume that the owner's outside option is zero.

I assume that effort $e_i$ is observable only by the $i$th worker and that performance $x_i$ is observable by the manager and workers. The owner knows only the expectation of $e_i$ and $x_i$. Total performance $y$ is observable to everyone. However, since effort and performance are not verifiable by a court of law, the owner and the manager can refuse to give appropriate bonuses.

I assume that all players are rational and this is a common knowledge. I use a multilateral contract scheme, which is forming a contract among the manager and all workers at once based on performances of all workers (see \cite{10.1162/003355302760193968}), Thus, the members of the contract are stationary. Instead, I assume that workers select the best effort level for themselves, and the manager designs an appropriate bonus mechanism. Rational workers will not select a low effort level under an appropriate bonus scheme.

The manager's salary and bonus can be included in the total team salary and bonus or assigned separately by the owner. I address the former case in Section \ref{sec2} and the latter case in Section \ref{sep}. The case where the owner does not hire a manager and distributes the bonus equally based on total team performance is handled in \cite{kvaloy2019relational} and reviewed in Subsection \ref{comparison}.

\section{Bonus distribution including the manager}\label{sec2}
In this section, I address the case where the owner employs a manager and delegates the authority of distribution of salary and bonus of the whole team, including the manager himself or herself. I consider a twofold contract as follows. The workers exert effort $e_i$ that leads to $\bold{x}=(x_1, \dots, x_n)$. The owner decides the total salary $\alpha_0$ and a total bonus $b_0(y)$ depending on the overall performance $y(=\sum{x_i})$ that he or she can freely observe. I denote this part as the \textit{outer contract}. Then, the manager decides the salary $\alpha_i$ and the bonus $b_i(\bold{x})$ depending on performance vector $\bold{x}$. The \textit{inner contract} is defined as the contract between the manager and the workers regarding the distribution of the total salary and bonus. Note that the salary payment is legally enforceable; thus, the owner and the manager cannot refuse to pay it.

In each period, the owner gives salary $\alpha_0$ to the manager, and the manager distributes salary $\alpha_i$ to the workers. (The net salary of the manager is $\alpha_m=\alpha_0-\sum_i\alpha_i$.) For simplicity, I assume that there are no constraints for $\alpha_i$ and $\alpha_0$, as in \cite{kvaloy2019relational}. Then, each worker selects and exerts effort $e_i$, which leads to performance $\bold{x}$. Next, the owner receives revenue $y=\sum{x_i}$ and gives bonus $b_0(y)(\geq0)$ to the manager. Finally, the manager gives a bonus $b_i(\bold{x})(\geq0)$ to each worker.

The one-period profit of the $i$th worker is $\alpha_i+b_i(\bold{x})-c(e_i)$. The manager earns $\alpha_0-{\sum_i\alpha_i}+b_0(y)-\sum_i{b_i(\bold{x})}$, and the owner earns $y-\alpha_0-b_0(y)$ in a period.

This case is interpreted as the situation that the manager and the workers can freely form side contract. Since the manager decides the bonus distribution in her discretion and claims the residual bonus, she can include any side transfer in the main relational contract if she want, Thus, there is no side contract in this case. Moreover, even if the manager receives separate bouns from the owner, if the side contract between the manager and the workers is available, the situation should be analyzed as this case.
\subsection{Solution}
To guarantee the existence of the solution and the validity of the first-order approach regarding $e_i$, which is widely discussed in the literature (\cite{kvaloy2019relational,10.1162/003355302760193968}), I assume the second-order conditions regarding $e_i$ with $b_i$ and $b_0$.
\begin{assumption}\label{soc}

${\partial^2 E(b_i)\over{\partial e_i}^2}$, ${\partial^2 E(b_0)\over{\partial e_i}^2}<c''(e_i)$.

\end{assumption}

In the main text, I only address the normal distribution. For simplicity, I define two functions regarding the normal distribution:

\begin{definition}
\begin{equation}
p_n\equiv {1\over n2^n}\lim_{w\to\infty}((erfc(-{w\over\sqrt2}))^nw-\int_0^w{(erfc(-{\bar{w}\over\sqrt{2}}))^nd\bar{w}}),
\end{equation}
\begin{equation}
\rho_n(\eta)\equiv {1\over n2^n}(\eta erfc({\eta\over\sqrt{2}})^n-\int_{0}^{\eta}{erfc({\tilde{w}\over\sqrt{2}})^nd\tilde{w}})
\end{equation}
\end{definition}
With the normal distribution centered of effort $e$ with variance $\sigma^2$ (same in the main text below), the solution can be given as follows:
\begin{proposition}\label{simplified_sol}
When the inner contract is feasible, at least one of the following statements holds:

1. Optimal effort $e^*$ is the only solution of ${\partial E(b_0)\over\partial e}=c'(e)$ and $c'(e^*)\leq{\delta\over{1-\delta}}(\alpha_0+E(b_0;e^*)-nc(e^*)
-\bar{u_0}[n]-n\bar{u}){p(n)\over\sigma}$.

2. The bonus mechanism follows a tournament with a threshold. A worker who achieves the maximum performance receives a bonus ${\delta\over{1-\delta}}(\alpha_0+E(b_0)-\sum{c(e_i)}-\bar{u_0}[n]-n\bar{u})$ if his or her performance exceeds $e^*$, and other workers do not receive any bonus. If none of the workers’ performances exceeds $e^*$, No one receives the bonus. Moreover, ${\partial E(b_0)\over\partial e_i}> c'(e^*)$.

Let $k_0\equiv\alpha_0+E(b_0)-\sum{c(e_i)}-\bar{u_0}[n]-n\bar{u}$, and $\bar{\bar{e}}$ be the solution of the following equation \eqref{noropte}:
\begin{equation}\label{noropte}
{{\sigma(1-\delta)}\over\delta}{c''(e^*)\over np_n}=(1-c'(e^*)).
\end{equation}
Then, when the outer contract is feasible, $k_0$ and the optimal solution $e^*$ satisfy the following:
\begin{equation}\label{e^*andk0_main}
c'(e^*)={\delta\over{1-\delta}}k_0{p(n)\over\sigma}.
\end{equation}
Moreover, at least one of the following statements holds.

1. The optimal solution $e^*=\bar{\bar{e}}$ with
\begin{equation}\label{norsubopte}
c'(e^*)\sigma(\sqrt{2\pi\over n}+{1\over n p_n})\leq{\delta\over{1-\delta}} (e^*-c(e^*)-\bar{u}-{\bar{u_0}[n]\over n}).
\end{equation}

2. $e^*$ satisfies
\begin{equation}
c'(e^*)\sigma(\sqrt{2\pi\over n}+{1\over n p_n})={\delta\over{1-\delta}} (e^*-c(e^*)-\bar{u}-{\bar{u_0}[n]\over n}).
\end{equation}

\end{proposition}
Proof of the proposition is given in the Appendix \ref{proof_simp_sol}.

\subsection{Solution structure}\label{str}
Fundamentally, I can separate the optimal effort of all relational incentive contracts into the global optimal case and global suboptimal (second-best) case. In the general multiparametric optimization theory \cite{PistikopoulosEfstratiosN.2021Moac}, no inequality constraints bind if we can achieve a globally optimal solution, and some inequality constraints bind if we cannot achieve the global optimal solution. Let me refer to the governing equation of the optimal effort (global optimality condition) as the \textit{optimal equation} and the binding inequality constraint(s) of the suboptimal effort (becoming the governing equation(s) of them) as \textit{subconstraint(s)}.

Both inner and outer contracts have only one subconstraint. Note that the inner contract can be used to analyze the problem where the owner can observe individual performance and there is no manager by setting $b_0(y)\equiv y$. In the inner contract, the optimal equation is ${\partial E(b_0)\over\partial e_i}|_{e_{-i}=e^*}=c'(e^*)$, and the subconstraint is $c'(e^*)\leq{\delta\over{1-\delta}}(\alpha_0+E(b_0;e^*)-nc(e^*)-\bar{u_0}[n]-n\bar{u}){p(n)\over\sigma}$. The optimal equation of the outer contract is \eqref{noropte}, and the subconstraint is $c'(e^*)\leq{\delta\over{1-\delta}} (E(y;e^*)-nc(e^*)-n\bar{u}-\bar{u_0}[n]-k_0){1\over\sqrt{2\pi n}}$.

The optimal equation of the inner contract does not include any information about $\delta, \bar{u}, \bar{u_0}[n]$ because the global optimal solution depends only on $b_0(y)$ and $c(e)$. In the inner contract, the surplus is maximized when the manager has complete bargaining power. Thus, the maximization of the manager's profit is the same as the maximization of the surplus. Moreover, the profit of workers from the contract is the same as their outside profit. In contrast, the optimal equation of the outer contract depends on $\delta$. According to Theorem \ref{outsol}, the total surplus is smaller than the first best for $\delta<1$ since the owner pursues his or her profit instead of the total surplus. This means the manager has rent over the owner for information about individual performances. The aspect that the owner does not have information about the performance of individual workers is the difference between the maximization of total surplus and the maximization of owner profit. The left-hand side of \eqref{noropte}, which represents the difference, depends on the other condition $\delta$. The main reason for the difference and the dependence can be explained intuitively as follows. If the equilibrium effort increases, by Assumption \ref{soc} and the convexity of the cost function, the bonus for the workers increases. Considering that the manager can terminate the contract by giving no bonus to the workers, the profit of the manager by the contract should be increased. Thus, the owner should yield more surplus to the manager, which is not beneficial to the owner.

The subconstraints of both the inner contract and outer contract are a function of $\delta$, the surplus, the cost function, and the probability density function. This is because the subconstraint arises from the endowment constraint and the first-order condition. The surplus is a source of benefit to the manager and the owner. Therefore, the maximum value of a feasible bonus is proportional to the surplus and $\delta\over{1-\delta}$. The cost function and the probability density function constitute the first-order conditions. Therefore, the form of subconstraints is $c'(e^*)\leq{\delta\over{1-\delta}}\times(surplus)\times(probability$ $of$ $winning$ $bonus)$.

\subsection{Profit analysis}
Based on Proposition \ref{simplified_sol}, the bonus and the profit of the owner and the manager can be calculated as follows:
\begin{corollary}\label{profits_simp}
The profit of the manager is determined as follows:
\begin{equation}\label{managerprofit}
k_0={(1-\delta)c'(e^*)\sigma\over\delta p(n)}.
\end{equation}
Additionally, if the subconstraint of the outer contract holds, the profit of the owner can be determined as follows:
\begin{equation}\label{ownerprofit}
{(1-\delta)c'(e^*)\sqrt{2\pi n}\sigma\over\delta}.
\end{equation}
\end{corollary}
Since both the owner and the manager can renege from giving a bonus, they need enough profit to prevent them from reneging. The profit they need to give the bonus is decided by the bonus they give multiplied by ${1-\delta}\over\delta$, which means the relative importance of the current benefit and one-period surplus. They give a bonus to induce effort. In the optimal contract, the marginal bonus-winning probability of the effort is $p(n)\over\sigma$ for the workers (the bonus is given by the manager) and $1\over\sqrt{2\pi n}$ for the team (the bonus is given by the owner). These are the effort-inducing abilities of the manager and the owner. Then, the manager needs to give bonus $c'(e^*)\sigma\over p(n)$ to induce effort $e^*$, and the owner needs to give bonus $c'(e^*)\sqrt{2\pi n}\sigma$. As a result, the bonuses are a function of $n$, $c(e)$, and $e^*$, and the profits are a function of $n$, $c(e)$, $e^*$, and $\delta$. (Note that they are also functions of $f$ in general.)

Because the profit of the workers is $0$, the ratio between the profits of the manager and the owner can be interpreted as the distribution of the surplus. The ratio of the profits of the owner and the manager can be obtained as Corollary \ref{dividesurplus}.
\begin{corollary}\label{dividesurplus}
When the subconstraint of the outer contract binds,
The owner and the manager divide the total surplus according to the proportion $p_n\sqrt{2\pi n}:1$.
\end{corollary}

The portion of the surplus taken by the owner increases until $n=5$ but decreases after $n=6$. Considering that the portion of the surplus taken by each party is inversely proportional to the marginal bonus-winning probability of effort for the bonus given by each party, this trend can be explained as the effect of three mechanisms. First, giving the bonus to the best performer produces an additional incentive compared to giving a bonus to workers who achieve performance above the threshold. This leads to the high effort-inducing ability of the manager when $n$ is small. Second, the expected bonus for a worker decreases as $1\over n$ with an increase over $n$. This is the main reason that the effort-inducing ability became small for high values of $n$. Third, since the performance distributions are independent, the variance of the total performance is $n$ times the variance of the performance of an individual worker. Then, the owner's effort-inducing ability decreases as $1\over\sqrt{n}$. These effects produce the trend of the optimal distribution of surplus. Therefore, the owner's share of the surplus is maximum at $n=5$ and decreases when $n$ further increases since the effort-inducing ability of the manager decreases faster than the ability of the owner. (This means that the manager must give a larger maximum bonus and needs more surplus to honor the contract.)

Note that when the performance follows a stationary distribution around the constant multiplication of the effort, the marginal bonus-winning probability for the effort is given as constant (it depends only on the shape of the distribution and the number of workers) for both the individual worker and the team. (see Corollary \ref{difference_ratio} in the appendix.) Then, since the level of effort induced by the manager and the owner is the same, and they also share the same discount factor, their need for profit is inversely proportional to the ability of effort inducing, which is the same as the marginal bonus-winning probability for the effort. Thus, the optimal ratio of the division of surplus between the manager and the owner is constant when the performance is distributed following a fixed distribution around a constant multiplication of effort.

\section{Separate bonus scheme for the manager}\label{sep}
In this section, I handle the case where the salary and the bonus of the manager are separately determined by the owner. In each period, the owner gives salary $\alpha_m$ to the manager and assigns total salary $\alpha_t$ and total bonus $b_t$ to the workers. In contrast to the former section, the total bonus can be legally bind, The point that there are no residual claimant of the remaining bonus allows the owner to commit the total worker bonus. Since the manager receives the separate bouns from the owner, she cannot treat the legally bind team bouns as her own bonus. The manager distributes salary $\alpha_i$ to the workers ($\sum_i\alpha_i\leq\alpha_t$). For simplicity, I assume that there are no other constraints for $\alpha_t$, $\alpha_i$, and $\alpha_m$, as in \cite{kvaloy2019relational}. Then, each worker selects and exerts effort $e_i$, which leads to performance $\bold{x}$. Next, the owner receives the revenue $y=\sum{x_i}$ and gives the bonus $b_m(y)(\geq0)$ to the manager. Finally, the manager gives the bonus $b_i(\bold{x})(\geq0)$ to each worker. (Likewise, $\sum_ib_i(\bold{x})\leq b_t$)
The remaining salary and bouns can assinged to parties or destroyed. By definition, they cannot be assigned to the workers. If they are assigned to the manager, the commitment power on the total worker bonus will disappear and the problem will be similar to the former section. (The only difference is the positivity of the net manager bonus and this constraint only lower the surplus. Thus, if the owner assigns the residual salary and bonus to the manager, then the contract will be outperformed by the former case.) If the remaining salary and bonus are returned to the owner, it can be reflected to the manager bouns and the problem becomes intractable without too stronus assumptions. Therefore, I assume that the remaining salary and bonus are destroyed. As in the former section, $e_i$ is observable only by the $i$th worker, and performance $x_i$ is observable by the manager and workers. The owner knows only the expectation of $e_i$ and $x_i$. Total performance $y$ is observable to everyone. Additionally, the owner and the manager can refuse to give appropriate bonuses.

In this case, the side contract is assumed to be unavailable. If it is available, the situation should follow the arguments in the former section. However, since the manager cannot take the residual bonus, he or she has the incentive to corrupt. Corruption can occur as follows. After the realization of performance, the manager secretly conspires with the worker, who should receive a minimal bonus. Then, the manager gives the whole bonus to the worker. Finally, the manager and the worker secretly divide the bonus. Since the side contract is assumed to be unavailable, this corruption is automatically considered as reneging, However, because of an external monitoring system, they can take only a portion $\phi$ of the bonus depending on the quality of the external monitoring system. (I assume that $\phi$ is given.) This point is dealt with as an additional constraint (see appendix).

The one-period profit of the $i$th worker is $\alpha_i+b_i(\bold{x})-c(e_i)$. The manager earns $\alpha_m+b_m(y)$, and the owner earns $y-\alpha_t-b_t$ in a period.

\subsection{Solution}
The solution for the separate bonus case can be obtained as follows (Note that $e^*$ refers to the optimal effort of the solution and can be different with the case in the former section):
\begin{proposition}\label{normalouter_sep}
Let
\begin{equation}\label{normal_out_eta}
\eta\equiv \sqrt{2}{erfc^{-1}(2-2(1-{nc(e^*)+n\bar{u}-\alpha_t\over b_t}-n({\phi(1-\delta)\over\delta}-{\alpha_m+E(b_m)-\bar{u_0}[n]\over b_t})^+)^{1\over n})^+}.
\end{equation}
Then, for the inner contract, $b_i(x_i)=b_t$ when $x_i>(e^*-\sigma\eta)$ and $x_i$ is maximal and $b_i(x_i)=0$ otherwise.
Moreover, $e^*$ satisfies
\begin{equation}\label{sol_normal_inner_sep}
c'(e^*)={b_t(p_n+\rho_n(\eta))\over\sigma}.
\end{equation}

The solution of the outer contract is given as follows, depending on $\phi$:

When $\phi=0$, the optimal $\bar{t}(e^*)$ is $-\infty$, and the first best effort is achieved. (i.e., $c'(e^*)=1$)

Otherwise, $\eta$ is given as the solution of the following:
\begin{equation}\label{opteta}
2^n(np_n-\eta{\phi(1-\delta)\over\delta})=\int_{0}^{\eta}{erfc({\tilde{w}\over\sqrt{2}})^nd\tilde{w}}.
\end{equation}

Moreover, the optimal effort $e^*$ is given as the solution of the following:
\begin{equation}\label{opteqn_sep_normal}
{\sigma\over\eta} c''(e^*)=1- c'(e^*).
\end{equation}
\end{proposition}

The solution of the inner contract, in this case, \eqref{sol_normal_inner_sep}, is a simple form in contrast to the solution of the previous case. This is because there is no need to separately consider the optimal solution without any constraints ($e$ satisfies the first part of Assumption \ref{soc}) in this case. Since the manager cannot take the remaining bonus, the manager needs only to maximize the performance. The cost of effort affects only the participation of the workers. Then, the manager seeks to make workers exert as much effort as possible unless they do not leave the contract.

\subsection{Solution structure}
In contrast to the previous case in which there is one optimal equation and one subconstraint for both the inner and outer contracts, in this case, the inner contract has only one subconstraint without any optimal equation, and the outer contract has only one optimal equation without any subconstraint. The manager's goal is to maximize total performance and thus maximize the equilibrium effort. Then, the optimal effort level without any constraints is infinity, and a valid optimal equation cannot exist. The inner contract has $c'(e^*)\leq {b_t(p_n+\rho_n(\eta))\over\sigma}$ as the subconstraint. Note that the subconstraint always holds for the optimal solution because there are no optimal equations. The outer contract does not have a subconstraint, which generally stems from the upper bound of the bonus imposed by the chance of reneging because the owner assigns the total bonus to the manager and cannot refuse to give it.

The subconstraint of the inner contract is a function of $\phi$, $b_t$, $\alpha_t$, $\bar{u}$, and the cost function $c(e)$. This is because the subconstraint arises from the bonus distribution constraint and the first-order condition. Although $e^*$ is the best threshold to induce maximal effort with fixed $b_t$, the manager can be forced to give the bonus with a lower threshold $e^*-\sigma\eta$ to prevent the workers from leaving by guaranteeing sufficient expected bonuses. Note that the optimal bonus for workers whose performance is lower than $e^*-\sigma\eta$ is still $0$. This means that lowering the threshold to earn a bonus is a better mechanism than giving a bit of a bonus to low performers. In contrast to the previous section, the subconstraint is independent of $\delta$ and $\bar{u_0}[n]$ when $\phi=0$ because there is no risk of reneging on the manager who cannot take the remaining team bonus.

The optimal equation of the outer contract does not depend on $\bar{u_0}[n]$ and does not depend on $\delta$ when $\phi=0$. The equation is determined only by the cost function $c(e)$. Since there is no possibility of reneging, the optimal effort is independent of the surplus and the discount factor.
\subsection{Bonus and Profit analysis}
Then, with Proposition \ref{normalouter_sep}, the team bonus $b_t$ and the profit of the manager and the owner can be calculated as follows (for a detailed explanation, see appendix):
\begin{corollary}\label{bonus_sep_final}
The equation describing the optimal total team bonus $b_t$ is given as follows:
\begin{equation}\label{optbt_sep_normal}
b_t={\sigma c'(e^*)\over p_n+\rho_n(\eta)}.
\end{equation}
Then, the destroyed surplus and the manager's profit are given as follows:
\begin{equation}
\sigma{erfc({\eta\over\sqrt{2}})^nc'(e^*)\over2^n(p_n+\rho_n(\eta))}, {\sigma(1-\delta)\over\delta}{\phi c'(e^*)\over(p_n+\rho_n(\eta))},
\end{equation}
respectively.

Finally, the owner's profit can be obtained as follows:
\begin{equation}\label{ownerprofit_sep}
ne^*-nc(e^*)-(\sigma{erfc({\eta\over\sqrt{2}})^n\over2^n}+{\sigma(1-\delta)\over\delta}\phi) {c'(e^*)\over(p_n+\rho_n(\eta))}-\bar{u_0}[n]-n\bar{u}.
\end{equation}
\end{corollary}

\section{Comparison of contracts}\label{sec4}

\subsection{Contracts without a manager}\label{withoutmanager}
In this subsection, I introduce relational contracts without the manager when the owner can or cannot observe individual performance.

To compare the contracts with the manager as a benchmark, I analyze the case where the owner can directly observe the individual performance of the workers and there is no manager. When I set $b_0(y)\equiv y$ and $\alpha_0\equiv\bar{u_0}[n]\equiv0$ in Proposition \ref{simplified_sol} to analyze the situation, the optimal equation of the inner contract becomes $c'(e^*)=1$. The subconstraint then becomes $c'(e^*)\leq{\delta\over({1-\delta})\sigma}(ne^*-nc(e^*)-n\bar{u}){p_n}$, which can be simplified as follows:
\begin{equation}\label{innernorsubopte}
{\sigma c'(e^*)\over n p_n}\leq{\delta\over{1-\delta}} (e^*-c(e^*)-\bar{u})
\end{equation}
Note that the constraint becoming more relaxed when $n$ goes up. This is because the more surplus generated and the owner can still observe individual performances.

Now, I consider the main situation of this paper that the owner cannot observe individual performances. In the former sections, the owner hires a manager since he or she cannot observe the performance of workers without the manager. However, the owner can give the same bonus to all workers based on total performance instead of hiring a manager. The monitoring, in this case, becomes less effective, but the inefficiency generated by the manager disappears.

In the direct equal compensation without a manager model, the owner gives $\alpha_i$ and $b_i(y)$ directly. This situation and observation are analyzed in Section \ref{sec2} of \cite{kvaloy2019relational}. I summarize it as follows.

Equation \eqref{EA} remains the same, and \eqref{EP} and \eqref{EM} are combined and replaced with ${\delta\over{1-\delta}}(E(y-\sum_ib_i(y))-\sum_i\alpha_i)\geq\sum_ib_i(y)$. Equation \eqref{IC} remains the same, and I still assume the validity of the first-order approach (Assumption \ref{soc}). Using the definition of $k_i$, I can express the optimization problem as follows: ($g(y;e_1,\dots,e_n)$ is the probability distribution of $y$ under the efforts.)
\begin{equation}
\max_{\alpha_i, b_i, e_i} E(y)-\sum_ik_i-\sum_{i}c(e_i)-n\bar{u}
\end{equation}
\begin{equation}
s.t. {\delta\over{1-\delta}}(E(y)-\sum_ik_i-\sum_{i}c(e_i)-n\bar{u})\geq\sum_ib_i(y)
\end{equation}
\begin{equation}
\forall i, \int{b_i(y)g_{e_i}(y;e_1,\dots,e_n)dy}-c'(e_i)=0
\end{equation}
\begin{equation}
\forall i, \forall y, b_i(y)\geq0
\end{equation}
\begin{equation}
\forall i, k_i\geq0
\end{equation}
Since $g(y,e_1,\dots,e_n)\propto N(\sum_i e_i,n\sigma^2)$, $g_{e_i}(y;e_1,\dots,e_n)$ is larger than $0$ if and only if $y\geq\sum_ie_i$ and $\int_{\sum_ie_i}^\infty{g_{e_i}(y;e_1,\dots,e_n)dy}={1\over\sigma\sqrt{2\pi n}}$.
By means of a similar procedure as that of Theorem \ref{T1}, I can easily obtain the fact that the optimal $k_i$ is $0$ for all $i$ and that this contract obtains that the first-best or the bonus of every worker is ${\delta\over{1-\delta}}(e^*-c(e^*)-\bar{u})$ if $y>ne^*$ and $0$ otherwise.

In summary, the optimal $e^*$ is the solution of $c'(e^*)=1$ if the solution satisfies $\sigma\sqrt{2\pi n}c'(e^*)\leq{\delta\over1-\delta}(e^*-c(e^*)-\bar{u})$. Otherwise, the optimal $e^*$ is the largest solution of $\sigma\sqrt{2\pi n}c'(e^*)={\delta\over1-\delta}(e^*-c(e^*)-\bar{u})$.

\subsection{Analytical Comparison}\label{comparison}
In this subsection, I compare the contracts by comparing the equations.

To compare the contracts well, some intuition must be developed regarding $\eta$. Considering that $erfc(x)<1$ for $\forall x>0$, Corollary \ref{Deltaeta} is a direct result of \eqref{opteta}.
\begin{corollary}\label{Deltaeta}
Let $\Delta\eta\equiv\eta-{np_n\delta\over\phi(1-\delta)}$, then $\Delta\eta>0$, and it goes to $0$ as $n$ increases.
\end{corollary}
Note that $\Delta\eta$ is typically a very small positive number. Then, $\eta$ is slightly larger than ${np_n\delta\over(1-\delta)}$ when $\phi=1$.

The optimal equation of two contracts without the manager is $c'(e^*)=1$, which means the first best. In these contracts, the owner does not need to distribute surplus and pursue the first best if possible. Conversely, in the other two contracts in which the manager is employed, the owner needs to give more surplus as he or she wants a higher effort level. Then, the owner is satisfied with a lower effort level than the first best. Substituting the definition of $\Delta\eta$ in \eqref{opteqn_sep_normal}, I obtain the following:
\begin{equation}\label{opteqn_sep_normal_Deltaeta}
{\sigma\over(\Delta\eta+{np_n\delta\over\phi(1-\delta)})} c''(e^*)=1- c'(e^*).
\end{equation}
In \eqref{noropte} and \eqref{opteqn_sep_normal_Deltaeta}, $\sigma\over np_n$ is the factor presenting the inverse of the effort-inducing ability by bonus the managers bonus, and $(1-\delta)\over\delta$ means the factor where the manager prefers current profit by reneging rather than potential future profit. In \eqref{opteqn_sep_normal_Deltaeta}, $\phi$ means the ease of the manager reneging. Note that the side payment is not allowed as the continuing relational contract in this seperate bonus scheme and the contract terminates if it occurs. The right-hand side of \eqref{opteta} or $\Delta\eta$ can be interpreted as the inefficiency of the inner contract of the case where the manager receives a separate salary and bonus from the owner, in which the manager inevitably gives bonuses to the best performer whose performance is below the optimal effort to prevent the workers from leaving.

Except for the case with the manager who receives separate income in which there is no subconstraint, the subconstraints of the three cases all have a similar structure:
\begin{equation}
\sigma H(n)c'(e^*)={\delta\over1-\delta}(e^*-c(e^*)-\bar{u}-({\bar{u_0}[n]\over n}))
\end{equation}
with $H(n)={1\over np_n}$, $\sqrt{2\pi n}$, and $\sqrt{2\pi\over n}+{1\over np_n}$ for the observable and unobservable individual performance (for the owner) without the manager case and team salary bonus including the manager case, respectively, and the last term is only for the case where the manager divides the team salary and bonus including himself or herself.

In these contracts, the owner and the manager need a surplus to ensure enough bonus to induce effort. The right-hand side presents the surplus multiplied by the importance of the current profit over the potential profit of future time steps. The left-hand side is the sum of all bonuses awarded by the owner and the manager needed to induce effort $e^*$. Considering $\forall n>1, {1\over np_n}<\sqrt{2\pi\over n}+{1\over np_n}<\sqrt{2\pi n}$, greater effort can be induced with the same surplus in the order of observable individual performance from the owner case, contract with the manager who divides income including himself or herself case, unobservable individual performance case and same bonus for workers case. In the observable effort in the owner case, giving the bonus to the best performer, provided that the performance is greater than $e^*$, is an efficient bonus mechanism. The additional term $\sqrt{2\pi\over n}$ in $H(n)$ with the manager represents the division of the surplus between the owner and the manager. It allows the lower effort to be induced with the same surplus. Giving the same bonus to all workers is an inefficient way to give a bonus since the total performance has a greater variance than individual performance. Note that $\sqrt{2\pi n}$ is the only increasing function of $n$ of these three $H(n)$s. This is the result of the inefficiency of the equal bonus scheme and makes the contract more inefficient when $n$ increases. Finally, note that ${\bar{u_0}[n]\over n}$ is the effect of the outside profit of the manager.

Then, I explain the meaning of each parameter and visualize the results by plotting the optimal effort, total surplus, $\bar{u_0}[n]$ to achieve a certain owner profit, and the benefit from the contract of parties as functions of $\sigma, n$, and $\phi$ in the case where $f(x_i;e_i)$ is a normal distribution with mean $e_i$ and variance $\sigma^2$ and the cost function $c(e)={1\over 2}e^2$. The probability distribution function and the cost function are sufficiently simple to compute and plot using MATLAB.

\subsection{Effects of uncertainty and market durability}
The standard deviation of the individual performance $\sigma$ indicates how the performance is affected by factors that the worker cannot control. It can be considered low when the work is simple and repetitive, such as handicraft manufacturing. Conversely, works affected severely by random factors or market environments such as dealing can be considered to have a high $\sigma$.

The discount factor $\delta$ can be an indicator of the durability of the market, the company, and employment. People affiliated with a firm with low employee turnover, a long period of existence, and a stable market might feel there is a high probability that the same situation will be repeated. This is what is indicated by $\sigma$ and $\delta$.

By Proposition \ref{simplified_sol}, Proposition \ref{normalouter_sep}, Corollary \ref{Deltaeta}, and the results of Subsection \ref{withoutmanager}, I can obtain the following property:
\begin{corollary}
All optimal equations, subconstraints, and manager profits can be expressed or approximated as functions of $\sigma(1-\delta)\over\delta$, $\bar{u}$, $\bar{u_0}[n]$, $f$, and $c(e)$.

The surplus and owner profits of all contracts can also be expressed as functions of them, except in the case where the manager receives a separate income. In that case, only the destroyed surplus term $\sigma{erfc({\eta\over\sqrt{2}})^n\over2^n}{c'(e^*)\over(p_n+\rho_n(\eta))}$ depends on $\sigma$ instead of $\sigma(1-\delta)\over\delta$.
\end{corollary}
Since $erfc(x)<1$ for $\forall x>0$, the destroyed surplus term is negligible. Then, because the change in $\delta$ can be treated as the equivalent change in $\sigma$, I present only the dependency on $\sigma$. As $\sigma$ increases from small to large values, the optimal effort follows the solution of the optimal equation until the subconstraint binds. When $\sigma$ is high, the left-hand side of the subconstraint is larger, and the subconstraint binds with optimal effort (second-best). For each contract with a subconstraint, there always exists a threshold value of $\sigma$ over which the subconstraint binds.

The analytical results for the effects of $\sigma$ and $\delta$ are summarized in Corollary \ref{sigmadelta}, which can be directly obtained from \eqref{noropte}, \eqref{norsubopte}, \eqref{opteqn_sep_normal}, and the results of the previous subsection.
\begin{corollary}\label{sigmadelta}
In contracts with a manager, the optimal effort and surplus are a decreasing function of $\sigma$ and an increasing function of $\delta$. In contracts without a manager, they remain constant (first best) when the optimal equation is valid.
\end{corollary}
\begin{figure}[h]
\centering
\begin{subfigure}[b]{0.49\textwidth}
\centering
\includegraphics[width=1.1\textwidth]{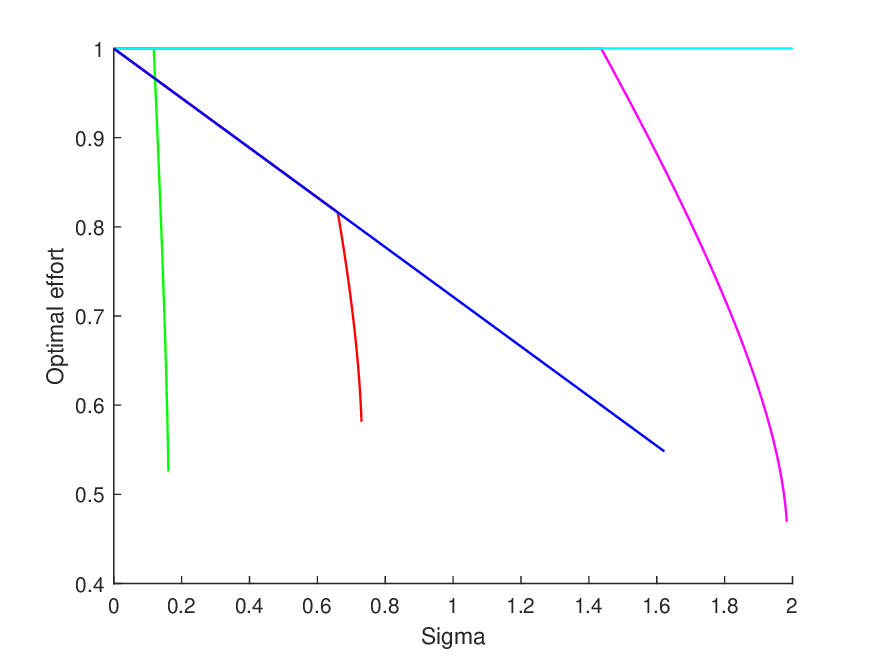}
\caption{Optimal effort ($n=10$)}
\end{subfigure}
\begin{subfigure}[b]{0.49\textwidth}
\centering
\includegraphics[width=1.1\textwidth]{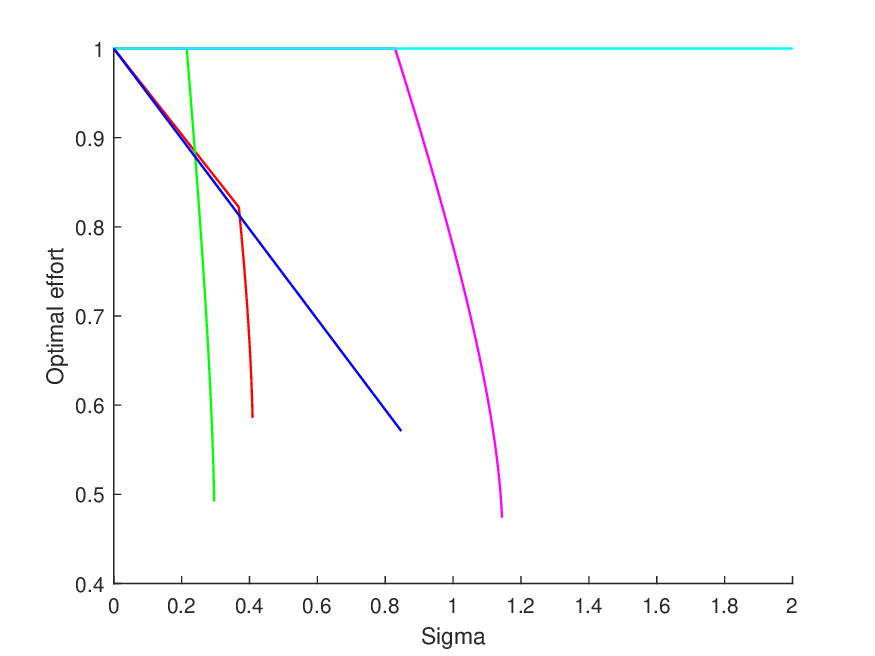}
\caption{Optimal effort ($n=3$)}
\end{subfigure}
\begin{subfigure}[b]{0.49\textwidth}
\centering
\includegraphics[width=1.1\textwidth]{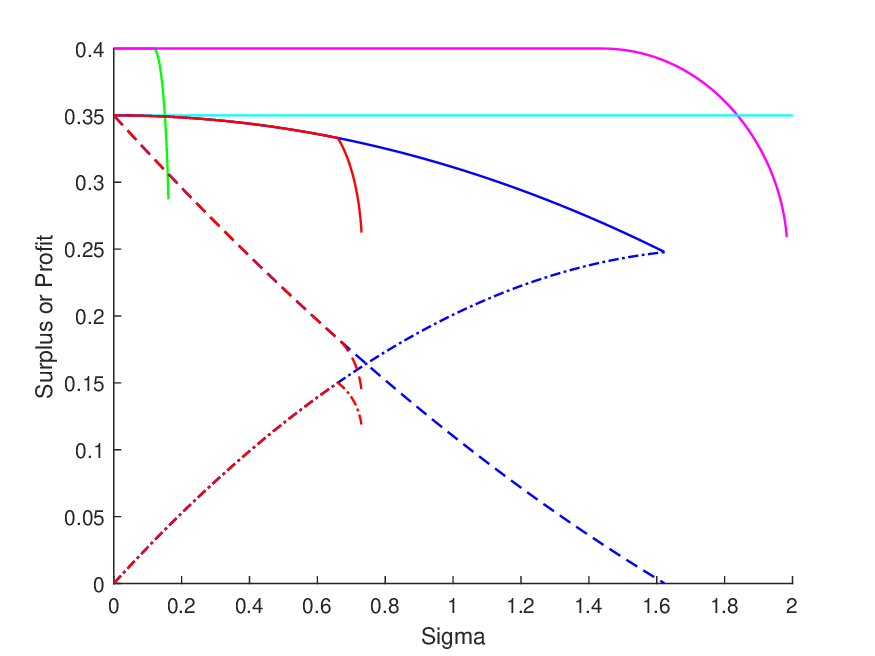}
\caption{Surplus and Profits per Worker ($n=10$)}
\end{subfigure}
\begin{subfigure}[b]{0.49\textwidth}
\centering
\includegraphics[width=1.1\textwidth]{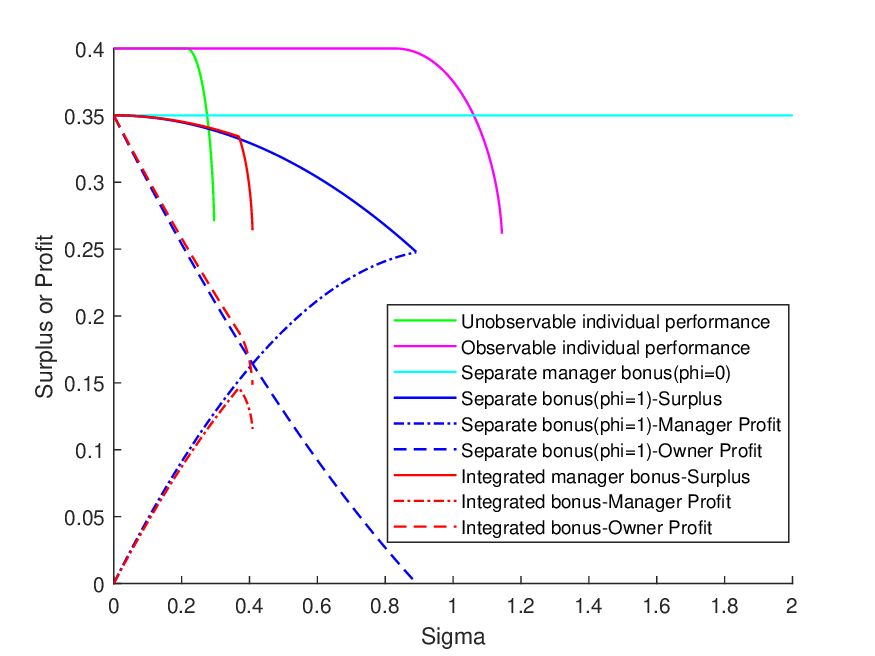}
\caption{Surplus and Profits per Worker ($n=3$)}
\end{subfigure}
\caption{Optimal effort, surplus and profits per worker of optimal contracts versus $\sigma$ ($\delta=0.7, \bar{u}=0.1, \bar{u_0}[n]=0.05n$)}
\label{discount_sigma_n}
\end{figure}

Figure \ref{discount_sigma_n} presents the optimal effort, surplus, and profits per worker of each party as a function of $\sigma$ when $\delta=0.7$, $n=3,10$. $\bar{u}$ and $\bar{u_0}[n]$ are set to $0.1$ and $0.05n$, respectively. Note that when there is no uncertainty in the performance, both the first-best surplus and the maximum owner profit are achieved. One can check that the optimal efforts are $1$ at $\sigma=0$ and decreasing functions of $\sigma$ at some point for all contracts except the separate bonus for the manager case with no possibility of corruption ($\phi=0$), in which the first best is always achieved. Surpluses and the owner profits are also the first best surplus when there is no uncertainty in the performance ($\sigma=0$) and decrease as the performance becomes more uncertain after some points for all cases except the separate bonus case. A larger number of workers allows high optimal efforts and surpluses to be achieved easily for all cases except the (no manager and) unobservable individual output (for the owner) case. This is because $1\over np_n$ and ${1\over np_n}+\sqrt{2\pi\over n}$ are decreasing functions of $n$, whereas $\sqrt{2\pi n}$ is an increasing function of $n$. (Please see \eqref{noropte}, \eqref{opteqn_sep_normal_Deltaeta}, and discussion about $H(n)$ in this subsection.)

In contracts with the manager (except in the no corruption case), the optimal effort, surplus, and owner profits start decreasing at $\sigma=0$, whereas the optimal effort remains $1$ for some range of low $\sigma$ when there is no manager. This is because the owner must distribute the surplus to the manager. The profit of the manager increases until the subconstraints bind. In a separate bonus case with corruption in which no subconstraint exists, the manager's profit continues to increase, and the contract is feasible until the owner's profit becomes $0$.

The integrated bonus contract outperforms the separate bonus contract with corruption when the optimal equation binds. This is because the separate bonus contract employs a bonus mechanism with a low threshold that is inefficient. Nevertheless, the difference becomes negligible when there are many workers, and the separate bonus contract does not have a subconstraint, which severely degrades the optimal effort under a high variance of performance.

The separate bonus contract with no corruption achieves first-best results because both the manager and the owner cannot renege. However, because of the presence of the manager outside income $\bar{u_0}[n]$, it is outperformed by the contract without the manager under low variance of performance.

Therefore, there are crossover points in which the best contract is changed. This will be addressed in-depth in the next subsection.

\subsection{Effects of the number of workers and the managing cost}
The number of workers $n$ in a team is a very important factor in the relational contract. To successfully monitor workers, the institution must employ a competent manager. The required competency and market income of the manager increases as the number of workers under his or her supervision increases. Moreover, the institution should provide materials and systems to help monitor the manager. The managing cost or the manager's outside income plus the supplemental material and system cost, $\bar{u_0}$, is a key factor that affects the feasibility of the contract and the owner's income.

The analytical results for the effects of $n$ and $\bar{u_0}[n]$ are summarized in Corollary \ref{nu0}, which can be directly obtained from \eqref{noropte}, \eqref{norsubopte}, \eqref{opteqn_sep_normal}, and the results of the previous subsection since $np(n)$ is an increase in $n$.
\begin{corollary}\label{nu0}
In all contracts addressed in this paper, the optimal equation is independent of $\bar{u_0}[n]$. In the contract with a manager who distributes the income, including himself or herself, the optimal effort when the subconstraint binds is a decreasing function of $\bar{u_0}[n]$. The surplus and the owner profits are always decreasing with the increase of $\bar{u_0}[n]$. In the contracts with the manager, the optimal effort, surplus per worker, and owner profit per worker are an increasing function of $n$ under constant $\bar{u_0}[n]$, and thus, greater $n$ can allow greater $\bar{u_0}[n]$ while achieving the same owner profit.
\end{corollary}
\begin{figure}[h]
\centering
\begin{subfigure}[b]{0.49\textwidth}
\centering
\includegraphics[width=1.1\textwidth]{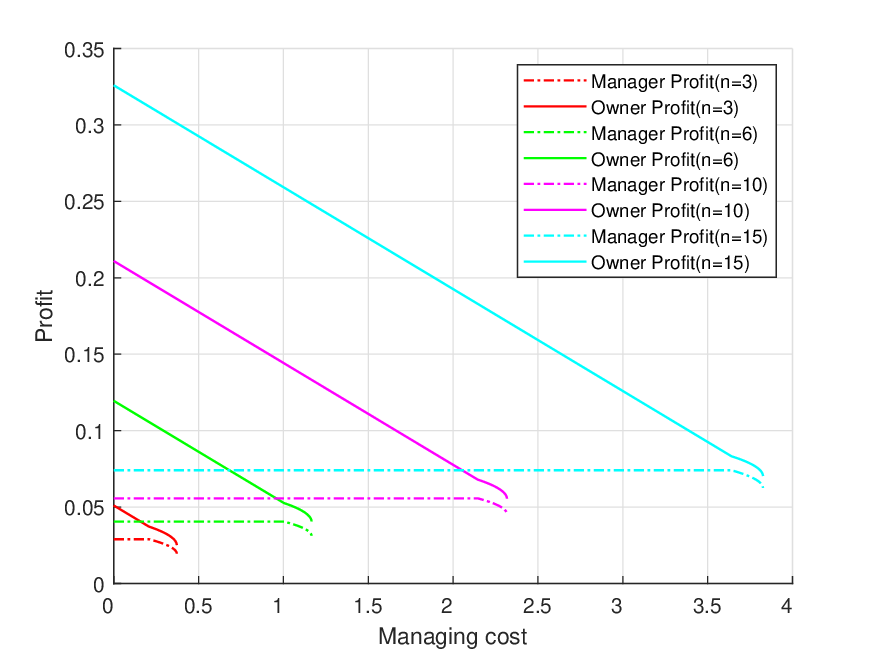}
\caption{Owner and manager profit per worker vs $\bar{u_0}$}
\label{profit_u0}
\end{subfigure}
\hfill
\begin{subfigure}[b]{0.49\textwidth}
\centering
\includegraphics[width=1.1\textwidth]{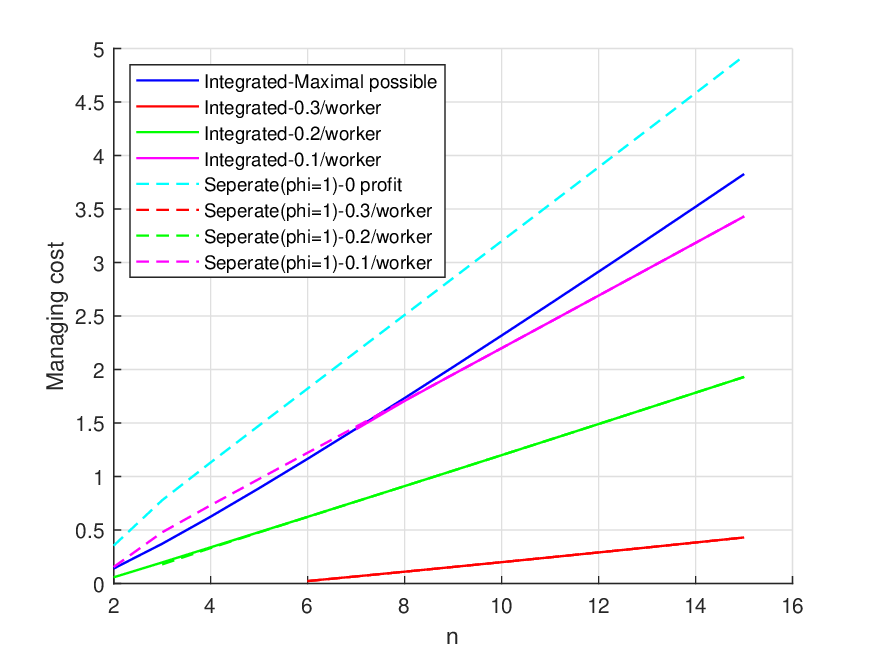}
\caption{Managing cost to achieve owner profit level}
\label{managing cost}
\end{subfigure}
\caption{Relation between $n$, $\bar{u_0}$, and profits ($\delta=0.7, \sigma=0.3, \bar{u}=0.1$)}
\end{figure}

In both results above, $\delta$ and $\sigma$ are set to $0.7$ and $0.3$, respectively. Figure \ref{profit_u0} presents the profit of the owner and the manager per worker in the case where the manager divides the bonus including himself or herself as a function of $\bar{u_0}[n]$ when $n=3,6,10,15$. The separate bonus case is not dealt with in Figure \ref{profit_u0} because the effect of $\phi$ will be presented in the next subsection, and the result of $\phi=1$ is almost the same as that in the integrated income case except for $n=3$. The manager's profit remains constant when the optimal equation binds and begins to decrease when the subconstraint binds. An increase in the managing cost makes the owner's profit decrease. When the subconstraint starts to bind, the slope of the owner's profit temporarily becomes gradual since the decrease in the surplus is divided by the decrease in the manager's profit. However, it becomes steep again when the managing cost further increases. Both profits increase when the number of workers increases. After the specific value of the managing cost depends (increases) on the number of workers, the contract becomes infeasible because there is no longer sufficient surplus to successfully induce effort.

Figure \ref{managing cost} presents the relation between the managing cost and the profit that the owner can achieve for each number of workers. In this figure, the manager can freely corrupt ($\phi=1$). For other values of $\phi$, see the next subsection. A higher managing cost is acceptable for a higher number of workers. The main reason for this effect is that the effort-inducing ability of the tournament with a threshold bonus mechanism increases by approximately $n$. The acceptable managing cost in the separate income case (dashed line) is slightly lower than that in the integrated income case with a small number of workers. Conversely, the feasibility of the integrated bonus contract is limited (blue line) by the subconstraint in the contract to a larger feasible region until the owner profit becomes $0$ (cyan dashed line). For instance, the separate income case can shape a feasible contract (magenta dashed line) in the upper region of the blue line.

\subsection{Effects of collusion (corruption of the manager)}
Although institutions make greater efforts to prevent improper collusion between their managers and workers, it is not always successful. The difficulty of collusion can be decided by the effectiveness of the internal audit system. I model how collusion is easy as $\phi$, the portion the accomplices can take through collusion compared to the money targeted by it.

The analytical results for the effects of $\phi$ are summarized in Corollary \ref{effectphi}, which is a direct result of Proposition \ref{normalouter_sep} and Corollary \ref{bonus_sep_final}.
\begin{corollary}\label{effectphi}
In all contracts addressed in this paper, the optimal effort, surplus, and owner profit are decreasing functions of $\phi$. The first best is achieved when $\phi=0$.
\end{corollary}

Since the effects of $\bar{u_0}$ only decrease the surplus and the owner profit for that amount, I set $\bar{u_0}[n]=0$ in the numerical results. The other parameters are set as $\delta=0.7$ and $\bar{u}=0.1$.
\begin{figure}[h]
\centering
\begin{subfigure}[b]{0.32\textwidth}
\centering
\includegraphics[width=1.1\textwidth]{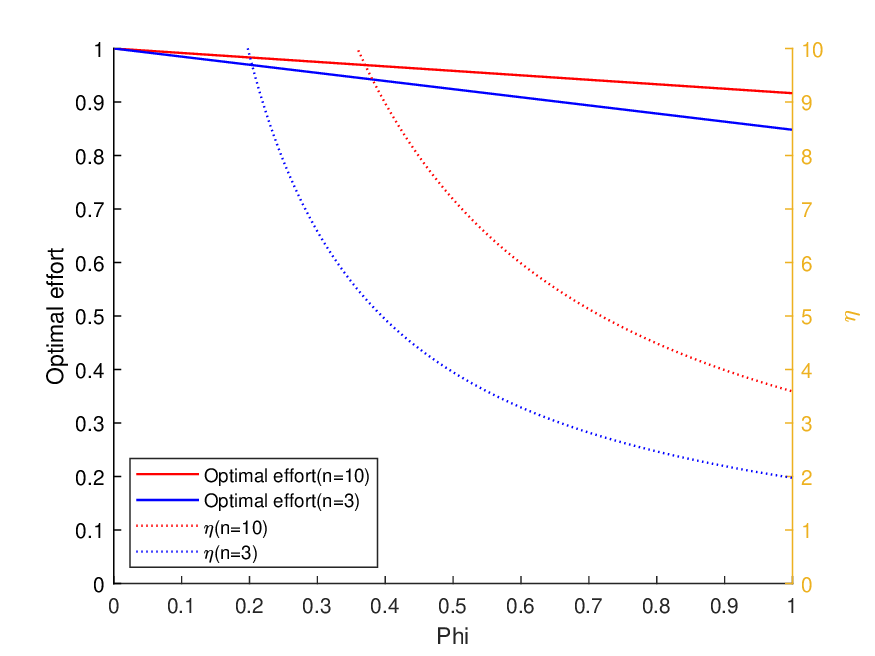}
\caption{Optimal effort and the threshold of giving the bonus vs. $\phi$}
\label{effort_eta_phi}
\end{subfigure}
\hfill
\begin{subfigure}[b]{0.32\textwidth}
\centering
\includegraphics[width=1.1\textwidth]{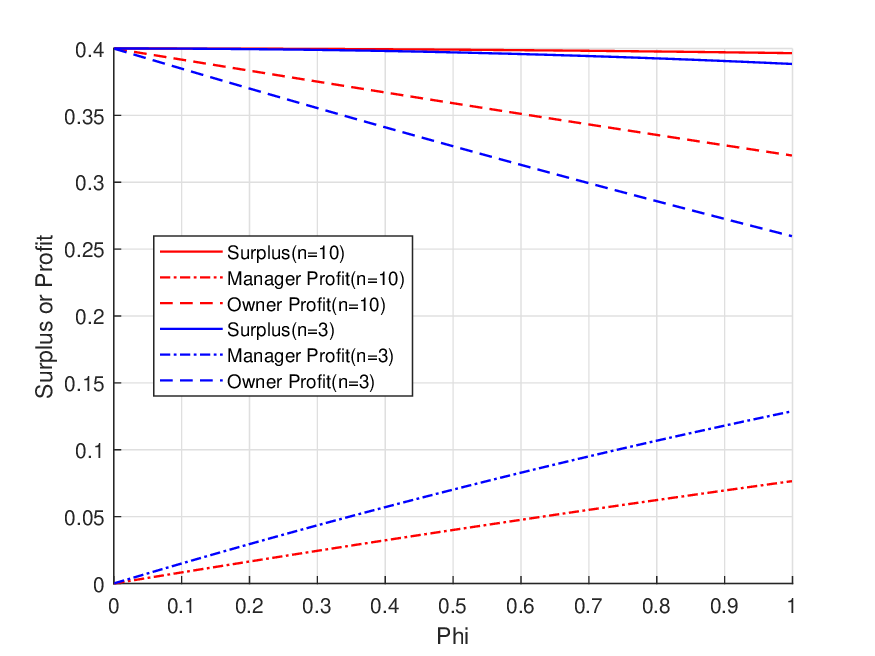}
\caption{Surplus, owner and manager profit per worker vs $\phi$($\sigma=0.3$)}
\label{surplus_phi_sigma=03}
\end{subfigure}
\hfill
\begin{subfigure}[b]{0.32\textwidth}
\centering
\includegraphics[width=1.1\textwidth]{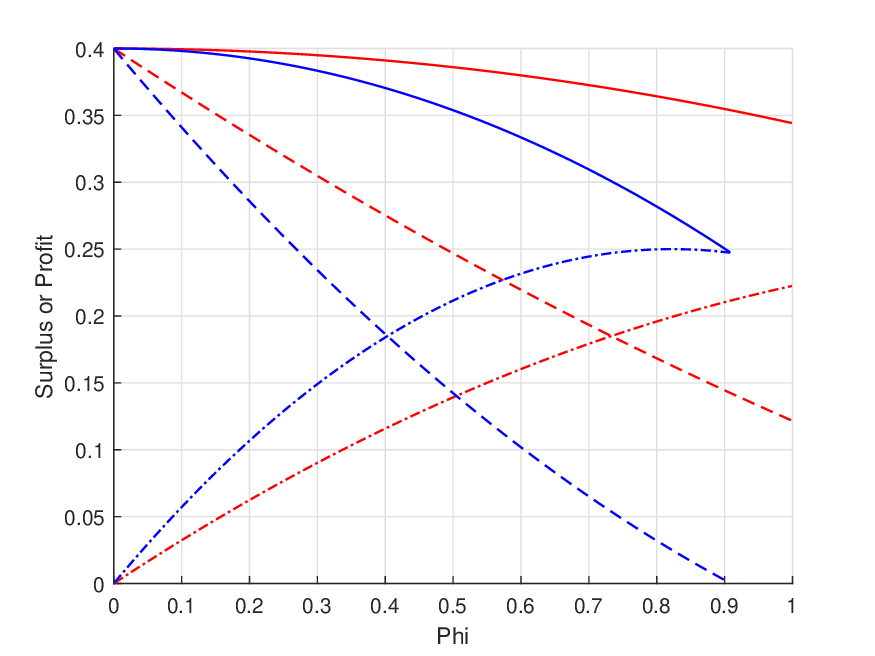}
\caption{Surplus, owner and manager profit per worker vs $\phi$($\sigma=1.2$)}
\label{surplus_phi_sigma=12}
\end{subfigure}
\caption{Effects of $\phi$ on the optimal effort, threshold of giving bonus, surplus per worker, and profits per worker ($\delta=0.7, \bar{u}=0.1,\bar{u_0}[n]=0$)}
\end{figure}
In Figure \ref{effort_eta_phi}, it is clear that the threshold of giving a bonus is very low, and high optimal effort is achievable when there is a good anti-corruption system ($\phi$ is small). The optimal effort and $\eta$ decrease when the availability of collusion increases. (Note that $\eta$ is the amount where the threshold is lower than the effort normalized by the standard deviation of the performance.) Easy collusion (high $\phi$) makes big profit have to be given to the manager to ensure the process of the same bonus, and it leads to low team bonus, a high threshold of giving a bonus to induce effort with a low total bonus, and a high salary. As a result, as in Figure \ref{surplus_phi_sigma=03} and Figure \ref{surplus_phi_sigma=12}, the surplus also decreases. Moreover, a higher possibility of manager corruption leads to a larger manager profit and reduces the owner's profit. When the owner's profit reaches below $0$, the contract becomes infeasible. These negative effects are intensified by a small number of team members and a large variance in performance. Since a higher managing cost ($\bar{u_0}[n]$) reduces only the surplus and the owner profit, the maximal managing cost makes the contract feasible in the same way as the owner profit of the point.

\section{Optimal management structure}
As the final part of the analysis, I address the decision of the owner on employing the manager. For simplicity, I assume that the managing cost $\bar{u_0}[n]>0$. Analytical results about the decision of the owner, who cannot observe the workers' performance, are summarized in Proposition \ref{unobserve_choice}.

\begin{proposition}\label{unobserve_choice}
The following statements hold:

1. When $\sigma$ is sufficiently small (smaller than some positive threshold), the best choice is to give the bonus equally based on total performance without employing a manager.

2. When $\bar{u_0}[n]\leq U_n(\phi)$, which is a decreasing function of $\phi$, employing a manager and giving a separate salary and bonus to the manager is the choice that can be feasible until the largest $\sigma$. When $\bar{u_0}[n]>U_n(\phi)$, giving the bonus equally based on total performance can be feasible until the largest $\sigma$.

3. When $\phi$ is sufficiently large and $\bar{u_0}[n]$ is sufficiently small, there exists a $\sigma$ that makes employing a manager and allowing him or her to distribute bonuses, including himself or herself, the best choice.
\end{proposition}

\begin{figure}[h]
\centering
\begin{subfigure}[b]{0.327\textwidth}
\centering
\includegraphics[width=1.1\textwidth]{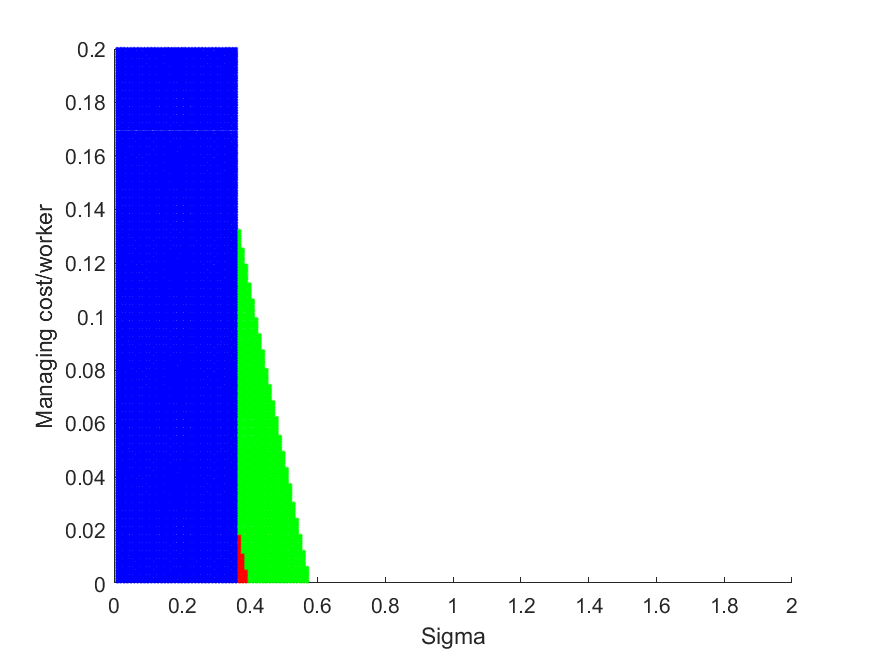}
\caption{Owner's best choice ($n=2$)}
\label{best_n=2_u0}
\end{subfigure}
\hfill
\begin{subfigure}[b]{0.327\textwidth}
\centering
\includegraphics[width=1.1\textwidth]{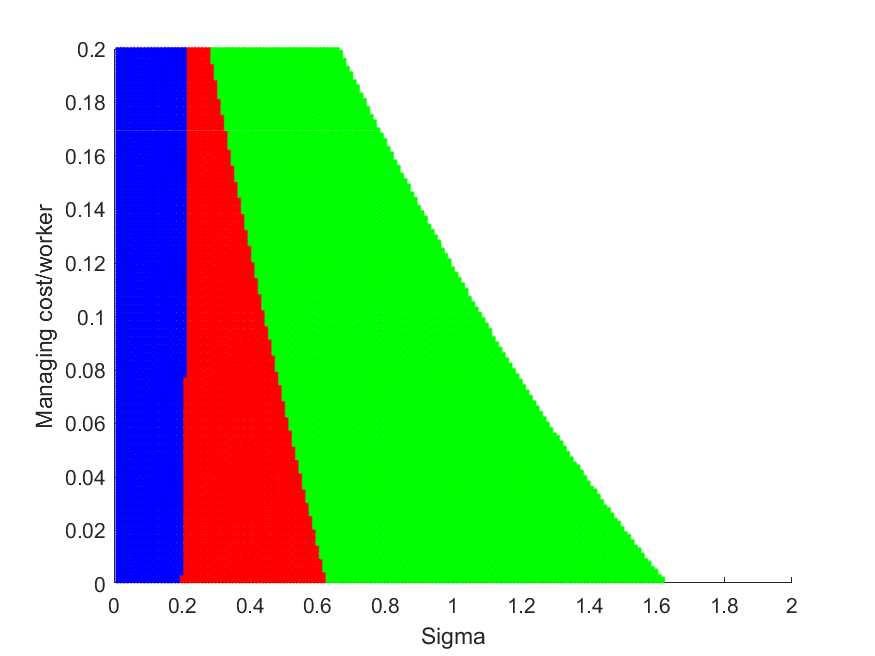}
\caption{Owner's best choice ($n=6$)}
\label{best_n=6_u0}
\end{subfigure}
\hfill
\begin{subfigure}[b]{0.330\textwidth}
\centering
\includegraphics[width=1.1\textwidth]{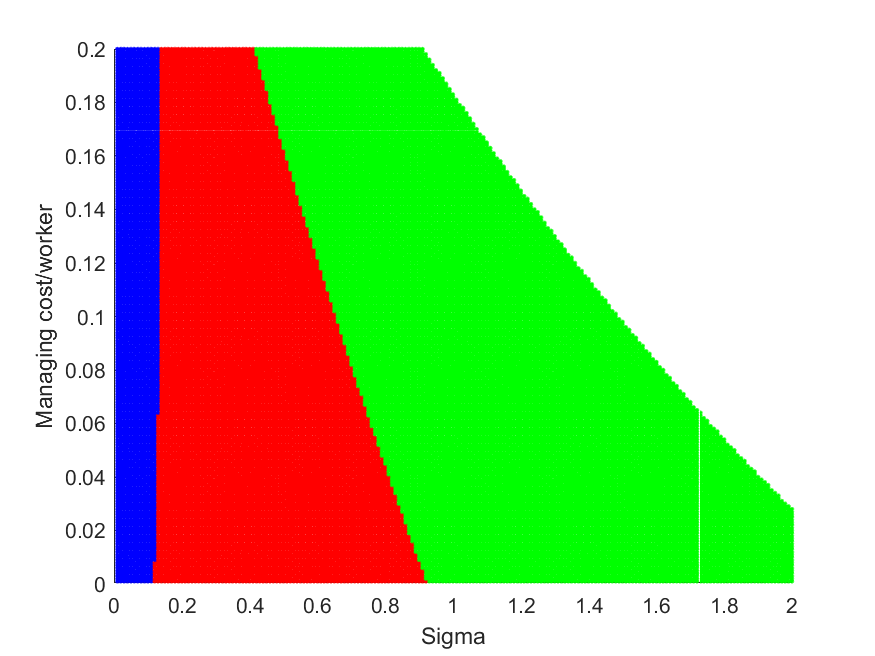}
\caption{Owner's best choice ($n=15$)}
\label{best_n=15_u0}
\end{subfigure}
\caption{Manager's best choice for various $n. \sigma, \bar{u_0}[n]$)($\delta=0.7, \bar{u}=0.1,\phi=1$)}
\label{choice_u0}
\end{figure}
The owner's best choice for various $n. \sigma$, and $\bar{u_0}[n]$ are presented in Figure \ref{choice_u0}. Blue, red, and green areas represent the best choice as giving equal bonuses without employing a manager, employing a manager with a team bonus including the manager bonus, and employing a manager with a separate bonus, respectively. Blank areas mean that no contracts are feasible. When giving an equal bonus without employing a manager is an optimal management structure, then $\sigma$ is small. The main reason for this is that the high discount factor allows the employer to give a sufficiently high bonus (without reneging) to induce effort even with equal distribution of team bonuses. This is an inefficient bonus schema and bonus mechanism with a manager and is the best choice for a high $\sigma$ because the owner can commit the bonus by assigning it to the manager. When $\bar{u_0}$ is low, there is some value of $\sigma$ that makes the integrated bonus with a manager optimal. This is because the separate bonus contract includes an inefficient bonus mechanism that has a low threshold. For a high $\bar{u_0}$, an integrated bonus scheme with a manager becomes infeasible for a smaller $\sigma$. When $n$ increases, the inefficiency of the equal bonus contract increases, and the effort-inducing ability of the tournament mechanism with a threshold strengthens. This leads to a condition such that giving equal bonuses is best for smaller areas, and contracts with managers are feasible for a higher $\sigma$.

\begin{figure}[h]
\centering
\begin{subfigure}[b]{0.32\textwidth}
\centering
\includegraphics[width=1.1\textwidth]{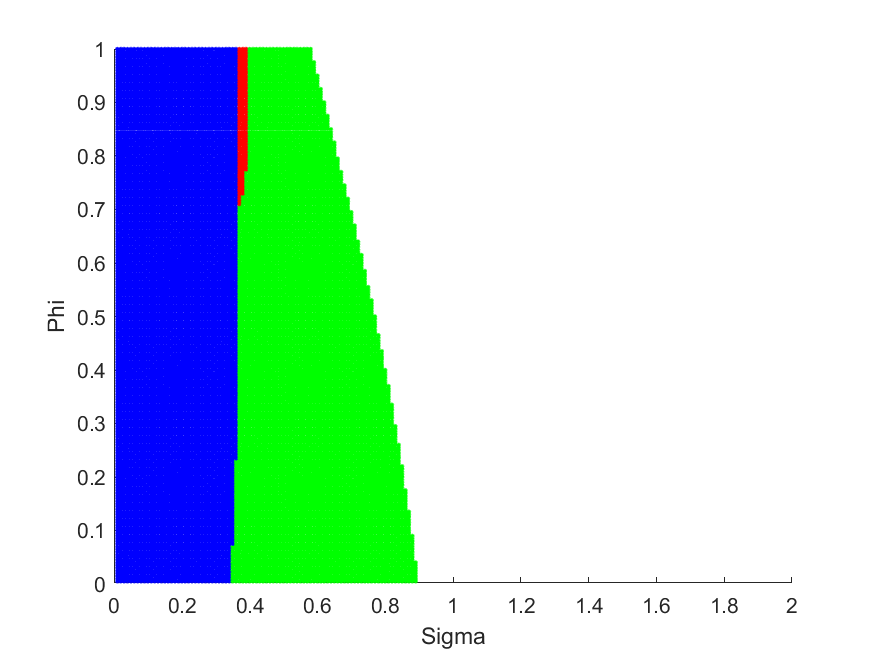}
\caption{Owner's best choice ($n=2$)}
\label{best_n=2_phi}
\end{subfigure}
\hfill
\begin{subfigure}[b]{0.32\textwidth}
\centering
\includegraphics[width=1.1\textwidth]{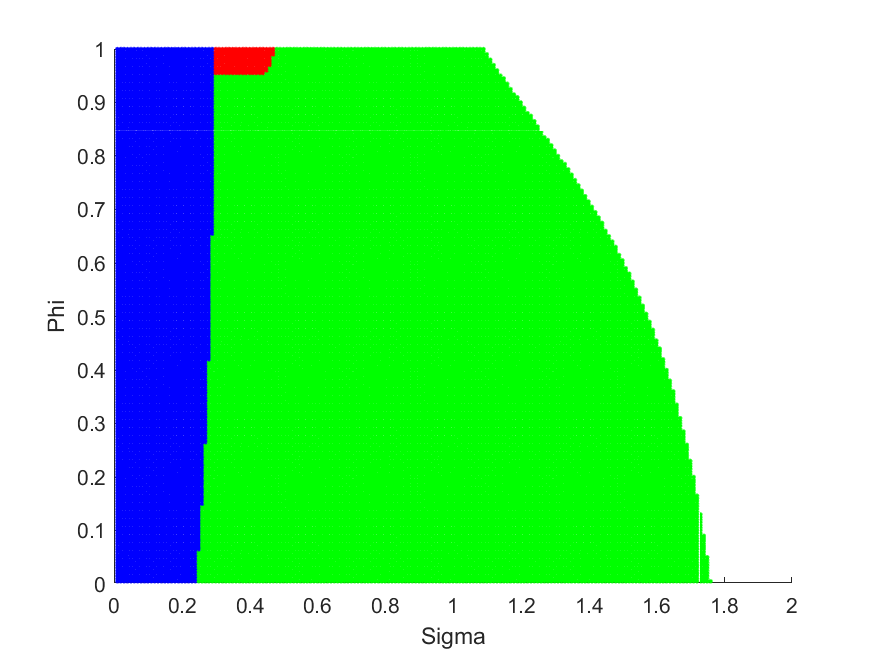}
\caption{Owner's best choice($n=3$)}
\label{best_n=3_phi}
\end{subfigure}
\hfill
\begin{subfigure}[b]{0.32\textwidth}
\centering
\includegraphics[width=1.1\textwidth]{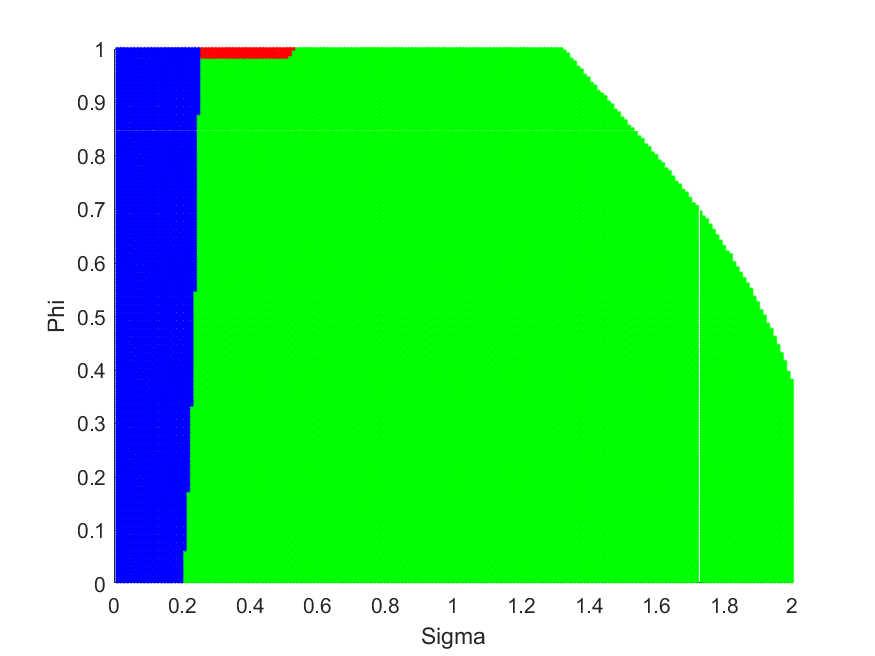}
\caption{Owner's best choice($n=4$)}
\label{best_n=4_phi}
\end{subfigure}
\caption{Manager's best choice for various $n. \sigma, \phi$)($\delta=0.7, \bar{u}=0.1,\bar{u_0}[n]=0$)}
\label{choice_phi}
\end{figure}
The owner's best choices for various $n. \sigma$ and $\phi$ are presented in Figure \ref{choice_phi}. When the corruption becomes harder ($\phi$ decreases away from $1$), giving a separate bonus to a manager begins to outperform giving a team bonus, including the manager's bonus. This is because the owner must give a smaller bonus to prevent reneging, and this effect overcomes the effect of the inefficient bonus scheme. When there are many workers, the inefficiency of the separate bonus contract becomes negligible, and giving a separate bonus to a manager becomes the optimal management structure, even when collusion between the manager and a worker is easy.

This result can also be partially helpful in understanding subcontracting in Britain up to 1870 \cite{10.2307/589888}. Since there were no cost management systems or elaborate internal audit systems in the past, it can be understood that the managing cost was small, and corruption could easily occur ($\phi$ is near $1$). Moreover, there were few assistants under the subcontract. Thus, such situations satisfied the conditions under which the contract with a manager and the distribution of the salary and bonus, including the manager himself and herself, is delegated, which is similar to a subcontract system, is the best option.
\section{Concluding Remarks}
I investigate when an owner who can observe only the total team performance hires a manager to supervise the workers. I show that the variance in the performance, the number of workers in a team, and the outside income of the manager are important determinants in the decision regarding employing a manager.

I characterize two stationary relational contracts among an owner, a manager, and workers. In one case, the manager's salary and bonus are included in the team's salary and bonus. In the other case, the manager's salary and bonus are given separately by the owner. I analyze them as twofold contracts. I consider the optimality of the inner contract between the manager and the workers as a constraint of the outer contract.

Then, I analyze the outer contract and derive the optimal equation and subconstraints of the outer contract. In both cases, the optimal team bonus scheme is a tournament with a threshold. The integrated bonus case has one optimal equation and one subconstraint in the outer contract. Conversely, the separate bonus case has one optimal equation and no subconstraints. The difference is due to the availability of bonus commitment.

I present the fact that the optimal contracts can be simplified with the normal probability distribution function. Moreover, I prove that the optimal portions of the surplus that the owner and manager take are constant under a normal distribution in the integrated bonus case. I compare the results with those in the case where there is no manager and the owner can observe the individual outputs and the case where there is no manager and the owner can observe only the overall output and divide the team bonus equally among all team members. I visualize the results with some simple inputs using MATLAB and present the effect and interpretation of parameters. Finally, I analyze the optimal management structure for the owner that depends on parameters. These results can be widely applied to many firms that hire managers to oversee workers because the assumption that the owner can only observe team performance widely holds.

In future work, the effects of the correlation between the performance of workers will be an important research question. Moreover, $\bar{u_0}[n]$ is defined as the sum of the outside income of the manager and the supplementary cost for the manager's work (e.g., the cost of purchasing evaluation tools). Additional analysis may be needed at this point. Finally, future research should also address the question of the tradeoff between reducing $\phi$ and the cost of an audit.

\begin{appendix}
\section{Detailed derivation of the optimal contract: Bonus distribution including the manager case}
\subsection{Constraints}
I use backward induction to analyze the contract. I first derive the optimal inner contract and use it as a constraint to obtain the optimal outer contract. Since the contract is stationary and infinitely repeated, I must consider only one time step for \eqref{IC} and \eqref{IC_M}. First, the manager and workers maximize their profits, and the workers select the effort level that will achieve the maximum profit in one period.
\begin{equation}\label{IC}
e_i=\argmax_{e_i'} (\alpha_i+E(b_i)-c(e_i'))
\end{equation}
Note that \eqref{IC} allows the manager to make workers exert positive effort rather than $0$.
The manager selects the bonus function $b_i(\bold{x})$ and wages $\alpha_i$ that maximize his or her profits. That is,
\begin{equation}\label{IC_M}
(\alpha_1, \dots, \alpha_n, b_1, \dots, b_n)=\argmax_{\alpha_1', \dots, \alpha_n',b_1', \dots, b_n'} (\alpha_0-{\sum_i\alpha_i'}+E(b_0-\sum_i{b_i'}))
\end{equation}

Second, any participant can terminate the contract if his or her profit from the contract is less than his or her profit without the contract. Since worker performance is ultimately not verifiable, the owner and manager can choose to give no bonuses. I assume that if a player does not honor the relational contract, the other player loses trust, and this breaks the contract (see discussions in \cite{10.1162/003355302760193968}). If the owner gives no bonuses to the manager for distribution, the whole contract is automatically terminated. Due to the positivity of the bonus, should the owner or the manager wish to terminate the contract, the best time is immediately before the distribution from the owner to the manager or from the manager to the workers. Therefore, for the relational contract to be sustainable, the owner should be able to obtain more potential profit if he or she gives the appropriate bonus to the manager.
\begin{equation}\label{EP}
\forall y, -b_0(y)+{\delta\over{1-\delta}} (E(y-b_0)-\alpha_0)\geq0
\end{equation}
The manager should be able to obtain more potential profit if he or she gives an appropriate bonus to the workers.
\begin{equation}\label{EM}
\forall \bold{x}, -\sum_i{b_i(\bold{x})}+{\delta\over{1-\delta}} (\alpha_0-{\sum_i\alpha_i}+E(b_0-\sum_i{b_i})-\bar{u_0}[n])\geq0
\end{equation}
Assuming \eqref{IC}, a worker should exert effort $e_i$ rather than effort $0$ even if he or she decides to leave the contract during the next period. Thus, workers will not choose other effort levels after they start a period. Therefore, workers continue the contract when they earn a profit greater than $\bar{u}$ from the contract.
\begin{equation}\label{EA}
\alpha_i+E(b_i)-c(e_i)\geq\bar{u}
\end{equation}
\subsection{Formulation and simplification}
Now I can construct the optimization problem of $b_i$. Below is the optimization problem, which I can write for the contract between the manager and the workers as follows.
\begin{subequations}\label{inner_ori}
\begin{align}
& \max_{\alpha_1', \dots, \alpha_n',b_1', \dots, b_n'} &&(\alpha_0-{\sum_i\alpha_i'}+E(b_0-\sum_i{b_i'}))\label{inner_oriobj}\\
&\text{subject to}&&\eqref{IC}, \eqref{EM}, \eqref{EA}\\
& &&\forall i,\forall \bold{x}, b_i(\bold{x})\geq0
\end{align}
\end{subequations}

Assumption \ref{soc} allows \eqref{IC} to be replaced by an equation. I can simplify \eqref{inner_ori} by uniting \eqref{EM} and \eqref{EA} as in \cite{kvaloy2019relational}. Now I introduce a lemma that replaces the free variable to simplify the optimization problem.
\begin{lemma}\label{simp}
Let $k_i\equiv \alpha_i+E(b_i)-c(e_i)-\bar{u}$ and $k_0\equiv \alpha_0+E(b_0)-\sum{c(e_i)}-\bar{u_0}[n]-n\bar{u}$; then, the optimal $\{k_i\}$ is $k_1=k_2=\dots=k_n=0$. Furthermore, under Assumption \ref{soc}, the optimization problem \eqref{inner_ori} can be reduced to \eqref{optbi}.
\begin{subequations}\label{optbi}
\begin{align}
& \max_{(b_1, \dots, b_n)} &&(\alpha_0-\sum_i c(e_i)-n\bar{u}+E(b_0))\\
&\text{subject to}&&\forall i, {\partial E(b_i)\over\partial e_i}=c'(e_i)\label{mu}\\
& &&{\delta\over{1-\delta}}(\alpha_0+E(b_0)-\sum{c(e_i)}-\bar{u_0}[n]-n\bar{u})\geq\sum_i{b_i(\bold{x})}\label{lambda}\\
& &&\forall i,\forall \bold{x}, b_i(\bold{x})\geq0\label{tau}
\end{align}
\end{subequations}
\end{lemma}
\begin{proof}
By Assumption \ref{soc}, \eqref{IC} can be replaced with
\begin{equation}
{\partial E(b_i)\over\partial e_i}=c'(e_i).
\end{equation}
Then, by substituting $k_i\equiv \alpha_i+E(b_i)-c(e_i)-\bar{u}$, I eliminate $\alpha_i$ with $k_i$ to simplify the problem. I can do this because the manager can change $k_i$ by changing $\alpha_i$ without affecting $e_i$. Then, \eqref{inner_oriobj}, \eqref{EM}, and \eqref{EA} are reduced as follows.
\begin{equation}
\max_{(k,b)} (\alpha_0-\sum_i k_i-\sum_i c(e_i)-n\bar{u}+E(b_0))
\end{equation}
\begin{equation}
\forall \bold{x}, -\sum_i{b_i(\bold{x})}+{\delta\over{1-\delta}} (\alpha_0+E(b_0)-\sum_ik_i-\sum_i{c(e_i)}-\bar{u_0}[n]-n\bar{u})\geq0
\end{equation}
\begin{equation}\label{newEA}
k_i\geq0
\end{equation}
Since the objective includes the substitution of $k_i$ and \eqref{newEA} is the only inequality that gives a lower bound of $k_i$, the optimal $\{k_i\}$ is $k_1=k_2=\dots=k_n=0$. As a result, the optimization problem reduces to \eqref{optbi}.
\end{proof}
Note that $k_i$ denotes the benefit of the $i$th worker by the contract. Lemma \ref{simp} means that the manager can obtain the whole surplus by giving the minimum salary that prevents workers from leaving the contract. Workers cannot take any profit even though the effort levels are hidden information.
\subsection{Optimal Contract}\label{sec3}
In this section, I solve the optimization problem of the inner contract \eqref{optbi} and formulate and solve the optimization problem of the outer contract step by step.
To easily address the optimization problem, I assume the monotonic likelihood ratio property (MLRP). The MLRP is used in several papers (\cite{kvaloy2019relational,10.1162/003355302760193968}).
\begin{assumption}\label{mlrp}
(Monotone Likelihood Ratio Property)

Let $f_{e_i}={\partial f\over \partial e_i}$ and $f_{e_i}(x_i;e_i)\over f(x_i;e_i)$ be monotonically increasing over $x_i$.

Moreover, let $g(y,e_1,\dots,e_n)=\int_{x_1+\dots+x_n=y}{f(x_1;e_1)\dots f(x_n;e_n)}$; then, $g_{e_i}(y;e_1,\dots,e_n)\over g(y;e_1,\dots,e_n)$ is monotonically increasing over $y$.
\end{assumption}

By Assumption \ref{mlrp}, I can let unique $t(e_i)$ denote $x_i$, which satisfies $f_{e_i}(x_i;e_i)=0$. Note that $f_{e_i}(x_i;e_i)f(\bold{x}_{i^{-}};\bold{e}_{i^{-}})={f_{e_i}(x_i;e_i)\over f(x_i;e_i)}\prod_j{f(x_j,e_j)}$
\subsection{Inner Contract}\label{inner}
This subsection analyzes the optimization problem for inner contract \eqref{optbi}. The main idea of this subsection is introduced in \cite{10.1162/003355302760193968}, and this subsection is also similar to sections 3 and 3.1 of \cite{kvaloy2019relational}. The most important difference between this work and previous works is that $y$ is replaced with $\alpha_0+b_0(y)$. This transforms the optimization problem into a parametric optimization problem with parameters $\alpha_0$ and $b_0(y)$. Thus, I separate cases according to $\alpha_0$ and $b_0$. Moreover, I want to use the solution of the inner contract \eqref{optbi} as a constraint of the outer contract. Thus, the main goal of this subsection and the first part of the next subsection is to find simple equations about the optimal effort $e^*$ to use as a constraint.

Let $\bar{e}(b_0)$ be the symmetric effort level that satisfies ${\partial E(b_0)\over\partial e_i}=c'(e_i)$. The solution is the same as $\bar{e}(b_0)$, which is the optimal solution without any constraints, or the bonus mechanism becomes a tournament with a threshold, and the upper bound of $b_i$ binds.

\begin{theorem}\label{T1}
When the reduced optimization problem is feasible, at least one of the following statements holds:

1. Optimal effort $e^*$ equals $\bar{e}(b_0)$ and $c'(e^*)\leq{\delta\over{1-\delta}}(\alpha_0+E(b_0;e^*)-nc(e^*)
-\bar{u_0}[n]-n\bar{u})\int_{t(e^*)}^\infty F(x_i,e^*)^{n-1}f_{e_i}(x_i;e^*)dx_i$.

2. The bonus mechanism follows a tournament with a threshold. A worker who achieves the maximum performance receives a bonus ${\delta\over{1-\delta}}(\alpha_0+E(b_0)-\sum{c(e^*)}-\bar{u_0}[n]-n\bar{u})$ if his or her performance exceeds $t(e^*)$, and other workers do not receive any bonus. Moreover, ${\partial E(b_0)\over\partial e^*}> c'(e^*)$.
\end{theorem}
\begin{proof}
We need $\alpha_0+E(b_0)-\sum{c(e_i)}-\bar{u_0}[n]-n\bar{u}\geq0$ because of \eqref{lambda} and \eqref{tau}. If $\alpha_0+E(b_0)-\sum{c(e_i)}-\bar{u_0}[n]-n\bar{u}\geq0$, $e_i=0, b_i\equiv0$ can be a feasible solution.

The Lagrangian of \eqref{optbi} can be written as follows. Let $\mu_i$, $\lambda(\bold{x})$ and $\tau_i(\bold{x})$ be Lagrange multipliers of \eqref{mu}, \eqref{lambda} and \eqref{tau}. $i^-$ denotes $1,\dots,i-1,i+1,\dots,n$.
\begin{equation}\label{lanbi}
\begin{split}
L_1=(\alpha_0-\sum_i c(e_i)-n\bar{u}+E(b_0))+\sum_i{\int{\tau_i(\bold{x})b_i(\bold{x})d\bold{x}}}+\\
\sum_i{\mu_i(\int{b_i(\bold{x})f_{e_i}(x_i;e_i)f(\bold{x}_{i^{-}};\bold{e}_{i^{-}})d\bold{x}}-c'(e_i))}+\\
\int{\lambda(\bold{x})({\delta\over{1-\delta}}(\alpha_0+E(b_0)-\sum_i{c(e_i)}-\bar{u_0}[n]-n\bar{u})-\sum_i{b_i(\bold{x})})d\bold{x}}
\end{split}
\end{equation}
\begin{equation}\label{lanbi}
\begin{split}
{\partial L_1\over \partial e_i}=(1+{\delta\over{1-\delta}}\int{\lambda(\bold{x})d\bold{x}})({\partial E(b_0)\over \partial e_i}-c'(e_i))+\sum_j{\mu_j{\partial ({\partial E(b_j)\over\partial e_j}-c'(e_j))\over \partial e_i}}
\end{split}
\end{equation}
Due to symmetry and Assumption \ref{soc}, I can assume all $\mu_i$ and optimal $e_i$ are the same over $i$.
Let $\mu_1=\mu_2=\dots=\mu_n\equiv\mu, e_1^*=e_2^*=\dots=e_n^*\equiv e^*$.
\begin{equation}\label{lanbi}
\begin{split}
{\partial L_1\over \partial b_i(\bold{x})}=\mu f_{e_i}(x_i;e_i)f(\bold{x}_{i^{-}};\bold{e}_{i^{-}})-\lambda(\bold{x})+\tau_i(\bold{x})
\end{split}
\end{equation}

If $\mu<0$, $\tau_i(\bold{x})>0$ and $b_i(x_i)=0$ when $x_i>t(e^*)$. $\lambda(\bold{x})=\mu\min{f_{e_i}(x_i;e_i)f(\bold{x}_{i^{-}};\bold{e}_{i^{-}})}$ if $\exists i, x_i<t(e^*)$. $b_i(x_i)={\delta\over{1-\delta}}(\alpha_0+E(b_0)-\sum{c(e_i)}-\bar{u_0}[n]-n\bar{u})\geq0$ when $x_i<t(e^*)$ and $x_i$ is minimal. Clearly, this case leads to ${\partial E(b_i)\over\partial e_i}<0$ and collides with Assumption \ref{soc}.

The next case I consider is $\mu>0$. $\tau_i(\bold{x})>0$ and $b_i(x_i)=0$ when $x_i<t(e^*)$.
$\lambda(\bold{x})=\mu\max{f_{e_i}(x_i;e_i)f(\bold{x}_{i^{-}};\bold{e}_{i^{-}})}$ if $\exists i, x_i>t(e^*)$. $b_i(x_i)={\delta\over{1-\delta}}(\alpha_0+E(b_0)-nc(e^*)-\bar{u_0}[n]-n\bar{u})\geq0$ when $x_i>t(e^*)$ and $x_i$ is maximal. Then, by $(1+{\delta\over{1-\delta}}\int{\lambda(\bold{x})d\bold{x}})({\partial E(b_0)\over \partial e_i}-c'(e_i))+\sum_j{\mu_j{\partial ({\partial E(b_j)\over\partial e_j}-c'(e_j))\over \partial e_i}}=0$, ${\partial E(b_0)\over \partial e_i}-c'(e_i)=-{\sum_j{\mu_j{\partial ({\partial E(b_j)\over\partial e_j}-c'(e_j))\over \partial e_i}}\over(1+{\delta\over{1-\delta}}\int{\lambda(\bold{x})d\bold{x}})}=-{\sum_j{\mu_j({\partial^2 E(b_i)\over{\partial e_i}^2}-c''(e_i))}\over(1+{\delta\over{1-\delta}}\int{\lambda(\bold{x})d\bold{x}})}>0$.

By ${\partial L_1\over \partial e_i}=0$, $\mu=0$ only if ${\partial E(b_0)\over \partial e^*}-c'(e^*)=0$. Thus, by Assumption \ref{soc}, $\mu=0$ only if $e^*=\bar{e}$. In this case, $c'(e^*)={\partial E(b_i)\over\partial e_i}$. To obtain the upper bound of ${\partial E(b_i)\over\partial e_i}$, I construct the following optimization problem.
\begin{subequations}\label{optexpbi}
\begin{align}
& \max_{(b_1, \dots, b_n)} &&\partial E(b_i)\over\partial e_i\\
&\text{subject to}&&{\delta\over{1-\delta}}(\alpha_0+E(b_0)-\sum{c(e_i)}-\bar{u_0}[n]-n\bar{u})\geq\sum_i{b_i(\bold{x})}\label{lambdatilde}\\
& &&\forall i,\forall \bold{x}, b_i(\bold{x})\geq0\label{tautilde}
\end{align}
\end{subequations}
Then, the Lagrangian of \eqref{optexpbi} can be obtained as follows. Let $\tilde{\lambda}(\bold{x})$ and $\tilde{\tau}_i(\bold{x})$ be Lagrange multipliers of \eqref{lambdatilde} and \eqref{tautilde}.
\begin{equation}
\tilde{L}_1={\partial E(b_i)\over\partial e_i}+\sum_i{\int{\tilde{\tau}_i(\bold{x})b_i(\bold{x})d\bold{x}}}+\int{\tilde{\lambda}(\bold{x})({\delta\over{1-\delta}}(\alpha_0+E(b_0)-\sum_i{c(e_i)}-\bar{u_0}[n]-n\bar{u})-\sum_i{b_i(\bold{x})})d\bold{x}}
\end{equation}
Now, we can obtain ${\partial \tilde{L}_1\over \partial b_i(\bold{x})}=f_{e_i}(x_i;e_i)f(\bold{x}_{i^{-}};\bold{e}_{i^{-}})-\tilde{\lambda}(\bold{x})+\tilde{\tau}_i(\bold{x})$ and (with the same reason with $\mu>0$ case) this means that the incentive structure same as with the case $\mu>0$ is optimal. Thus, $c'(e^*)={\partial E(b_i)\over\partial e_i}\leq {\delta\over{1-\delta}}(\alpha_0+E(b_0;e^*)-nc(e^*)
-\bar{u_0}[n]-n\bar{u})\int_{t(e^*)}^\infty F(x_i,e^*)^{n-1}f_{e_i}(x_i;e)dx_i$
\end{proof}

I neglect the case where two or more workers achieve the same maximum performance over $t(e^*)$ because the case has a zero Lebesgue measure since $f$ is continuous. Theorem \ref{T1} states that we can achieve $e^*=\bar{e}$, or the upper bound of $b_i$ binds. Many papers that analyze only the inner contract have concluded at this step (\cite{kvaloy2019relational,10.1162/003355302760193968}). However, I want to use the solution of the inner contract as a constraint in the outer contract. Therefore, I need to obtain a clear equation for optimal effort.

\subsection{Outer Contract}\label{outer}
Now, I derive the optimal $\alpha_0$ and $b_0$ to analyze the whole contract. Due to the backward induction scheme, the result of the inner contract should be used as a condition. The objective of the optimization, \eqref{popt}, aims to maximize the profit of the owner. The optimization problem can be summarized as follows.
\begin{subequations}\label{popt}
\begin{align}
& \max_{(\alpha_0, b_0(y))} &&E(y-b_0(y))-\alpha_0\label{poptobj}\\
&\text{subject to}&&\forall y, -b_0(y)+{\delta\over{1-\delta}} (E(y-b_0)-\alpha_0)\geq0\label{b0max}\\
& &&
{\delta\over{1-\delta}}(\alpha_0+E(b_0;e)-nc(e)
-\bar{u_0}[n]-n\bar{u})\int_{t(e)}^\infty F(x_i,e)^{n-1}f_{e_i}(x_i;e)dx_i-c'(e)\geq0\label{tilde}\\
& &&{\partial E(b_0)\over\partial e_i}(e)-c'(e)\geq0\label{bar}\\
& &&\eqref{tilde}\times\eqref{bar}=0\label{min}\\
& &&\forall y, b_0(y)\geq0\label{b0min}\\
& && \alpha_0+E(b_0;e)-nc(e)-\bar{u_0}[n]-n\bar{u}\geq0\label{k0min}
\end{align}
\end{subequations}

Constructing the Lagrangian with constraints \eqref{tilde}, \eqref{bar}, and \eqref{min} will lead to extremely complicated mathematical calculations. Therefore, to solve \eqref{popt}, I need to simplify them first. To simplify the constraints, I remove the complementary slackness, which is difficult to address in optimization problems, by proving that we can achieve an equal or better solution with ${\delta\over{1-\delta}}(\alpha_0+E(b_0;e)-nc(e)
-\bar{u_0}[n]-n\bar{u})\int_{t(e)}^\infty F(x_i,e)^{n-1}f_{e_i}(x_i;e)dx_i-c'(e)=0$. Lemma \ref{tildetostar} characterizes the alternative solution that makes \eqref{tilde} hold and is better than or equal to the original solution.
\begin{lemma}\label{tildetostar}
If we have a set of $(b_0, \alpha_0, e^*)$ that satisfies \eqref{b0max}, \eqref{bar}, \eqref{min}, \eqref{b0min}, and \eqref{k0min}, and makes ${\delta\over{1-\delta}}(\alpha_0+E(b_0;e)-nc(e)
-\bar{u_0}[n]-n\bar{u})\int_{t(e)}^\infty F(x_i,e)^{n-1}f_{e_i}(x_i;e)dx_i-c'(e)>0$, then we can find another set that satisfies ${\delta\over{1-\delta}}(\alpha_0+E(b_0;e)-nc(e)
-\bar{u_0}[n]-n\bar{u})\int_{t(e)}^\infty F(x_i,e)^{n-1}f_{e_i}(x_i;e)dx_i-c'(e)=0$ and results in a not-smaller value of \eqref{poptobj} compared to original set $(b_0, \alpha_0, e^*)$.
\end{lemma}
\begin{proof}
Let $(b_0, \alpha_0, e^*)$ be a feasible solution. I can set a new $\alpha_0$ (equal to or less than the original $\alpha_0$) as follows.
\begin{equation}
\begin{split}
\bar{\alpha_0}\equiv{{1-\delta}\over{\delta}}{c'(e^*)\over\int_{t(e^*)}^\infty F(x_i,e^*)^{n-1}f_{e_i}(x_i;e^*)dx_i}-E(b_0;e^*)+nc(e^*)+\bar{u_0}[n]+n\bar{u}
\end{split}
\end{equation}
\eqref{k0min} holds because $c'(e^*)\geq0$ and $\int_{t(e^*)}^\infty F(x_i,e^*)^{n-1}f_{e_i}(x_i;e^*)dx_i\geq0$ by Assumption \ref{mlrp} and the definition of the $t$ function.
Then, I can obtain the following property.
\begin{equation}
\begin{split}
c'(e^*)={\delta\over{1-\delta}}(\bar{\alpha_0}+E(b_0;e^*)-nc(e^*)-\bar{u_0}[n]-n\bar{u})\int_{t(e^*)}^\infty F(x_i,e^*)^{n-1}f_{e_i}(x_i;e^*)dx_i
\end{split}
\end{equation}
Clearly, $(b_0, \bar{\alpha_0}, e^*)$ is feasible and not worse than $(b_0, \alpha_0, e^*)$ for problem \eqref{popt}.

\end{proof}
Lemma \ref{tildetostar} implies that there is an optimal solution of the outer contract that makes \eqref{tilde} hold. In other words, the optimal bonus scheme can always be achieved. I can intuitively explain this fact as follows. In the outer contract, we can select an appropriate $b_0(y)$ to adjust the conditions of the inner contract. Considering that we need to satisfy both inequalities (\eqref{bar} and \eqref{tilde}) and generally that a larger $b_0(y)$ is needed to satisfy \eqref{tilde} with a larger $e^*$, it is optimal to adjust $b_0(y)$ to make \eqref{tilde} hold.

Then, I can solve the optimization problem \eqref{popt}. The theorem below is the main solution of the outer contract. The optimal solution is twofold, similar to that for the inner contract. I can achieve a globally optimal solution or the upper bound of $b_0$ binds.
\begin{theorem}\label{outsol}
Let $\tilde{t}(e)$ satisfy ${g_{e_i}(\tilde{t}(e);e,\dots,e)\over g(\tilde{t}(e);e,\dots,e)}=0$ (uniquely defined by the second part of Assumption \ref{mlrp}) and $\bar{\bar{e}}$ be the solution of the following equation \eqref{genopte}.
\begin{equation}\label{genopte}
\begin{split}
{{(1-\delta)}\over\delta}{\partial\over\partial e^*}{c'(e^*)\over\int_{t(e^*)}^\infty F(x_i,e^*)^{n-1}f_{e_i}(x_i;e^*)dx_i}=n(\int{f_{e_i}(x_i,e^*)x_idx_i}-c'(e^*))
\end{split}
\end{equation}
Then, $k_0$ and the optimal solution $e^*$ satisfy
\begin{equation}\label{e^*andk0}
c'(e^*)={\delta\over{1-\delta}}k_0\int_{t(e^*)}^\infty F(x_i,e^*)^{n-1}f_{e_i}(x_i;e^*)dx_i.
\end{equation}
Moreover, at least one of the following statements holds.

1. The optimal solution $e^*=\bar{\bar{e}}$ with $c'(\bar{\bar{e}})\leq{\delta\over{1-\delta}} (E(y;e^*)-nc(e^*)-n\bar{u}-\bar{u_0}[n]-k_0)\int_{\tilde{t}(\bar{\bar{e}})}^\infty{g_{e_i}(y,\bar{\bar{e}})dy}$.

2. $e^*$ satisfies $c'(e^*)={\delta\over{1-\delta}} (E(y;e^*)-nc(e^*)-n\bar{u}-\bar{u_0}[n]-k_0)\int_{\tilde{t}(e^*)}^\infty{g_{e_i}(y,e^*)dy}$.
\end{theorem}
\begin{proof}
I replace \eqref{tilde} and \eqref{min} with the following equality constraint.
\begin{equation}\label{estar}
\begin{split}
c'(e^*)={\delta\over{1-\delta}}(\alpha_0+E(b_0;e^*)-nc(e^*)\\
-\bar{u_0}[n]-n\bar{u})\int_{t(e^*)}^\infty F(x_i,e^*)^{n-1}f_{e_i}(x_i;e^*)dx_i
\end{split}
\end{equation}
Ultimately, the analysis of the inner contract is summarized as \eqref{estar} and \eqref{bar}. Now, I can use these results freely as constraints of the optimization problem of the outer contract.

In addition, I can further simplify the constraints by replacing $\alpha_0$ with $k_0$.
\begin{equation}\label{s_y}
\forall y, -b_0(y)+{\delta\over{1-\delta}} (E(y)-nc(e^*)-n\bar{u}-\bar{u_0}[n]-k_0)\geq0
\end{equation}
\begin{equation}\label{t}
c'(e^*)={\delta\over{1-\delta}}k_0\int_{t(e^*)}^\infty F(x_i,e^*)^{n-1}f_{e_i}(x_i;e^*)dx_i
\end{equation}
\begin{equation}\label{u}
{\partial E(b_0)\over\partial e_i}(e^*)-c'(e^*)\geq0
\end{equation}
\begin{equation}\label{v_y}
\forall y, b_0(y)\geq0
\end{equation}
\begin{equation}\label{w}
k_0\geq0
\end{equation}
I can remove \eqref{w} because $k_0\geq0$ according to \eqref{t}.

Now, I solve \eqref{popt} with constraints \eqref{s_y}, \eqref{t}, \eqref{u}, \eqref{v_y}, and \eqref{w}.
Let $s(y), q, u, v(y)$ be Lagrange multipliers of \eqref{s_y}, \eqref{t}, \eqref{u}, and \eqref{v_y} and let $g(y;e)$ denote the probability density function of $y$. Then, the Lagrangian of \eqref{popt} can be written as follows.
\begin{equation}
\begin{split}
L_2=E(y;e^*)-nc(e^*)-n\bar{u}-\bar{u_0}[n]-k_0\\
+\int{(-b_0(y)+{\delta\over{1-\delta}} (E(y)-nc(e^*)-n\bar{u}-\bar{u_0}[n]-k_0))s(y)dy}\\
+q(c'(e^*)-{\delta\over{1-\delta}}k_0\int_{t(e^*)}^\infty F(x_i,e^*)^{n-1}f_{e_i}(x_i;e^*)dx_i)\\
+u(\int{b_0(y)g_{e_i}(y;e^*)dy}-c'(e^*))+\int{b_0(y)v(y)dy}
\end{split}
\end{equation}
\begin{equation}
{\partial L_2\over \partial b_0(y)}=-s(y)+ug_{e_i}(y,e^*)+v(y)
\end{equation}
\begin{equation}\label{roundk0}
{\partial L_2\over \partial k_0}=-1+\int{s(y)dy}-q{\delta\over{1-\delta}}\int_{t(e^*)}^\infty F(x_i,e^*)^{n-1}f_{e_i}(x_i;e^*)dx_i
\end{equation}
Let $g(y,e)\equiv g(y,e,e,\dots,e)$. By Assumption \ref{mlrp} $g_{e_i}(y;e)$ is monotonically increasing in $y$, and I can let $g_{e_i}(\tilde{t}(e);e)=0$. I separate the problem into two cases regarding $u$.
First, I consider the case of $u>0$. When $u>0$, $s(y)>0$ for $y>\tilde{t}(e^*)$ and $v(y)>0$ for $y<\tilde{t}(e^*)$. Therefore, $b_0(y)=0$ for $y<\tilde{t}(e^*)$ and $b_0(y)={\delta\over{1-\delta}} (E(y;e^*)-nc(e^*)-n\bar{u}-\bar{u_0}[n]-k_0)$ for $y>\tilde{t}(e^*)$.

Second, I consider the case of $u=0$. If $\exists y, s(y)>0$ then $v(y)=s(y)>0$ and ${\delta\over{1-\delta}} (E(y;e^*)-nc(e^*)-n\bar{u}-\bar{u_0}[n]-k_0)=b_0(y)=0$ by complementary slackness. This result leads to $\forall y, b_0(y)=0$ and $e^*=0$ by \eqref{u}, which means no contract. Thus, $\forall y,s(y)=0$ and $\int{s(y)dy}=0$. Then, $q=-{(1-\delta)\over\delta\int_{t(e^*)}^\infty F(x_i,e^*)^{n-1}f_{e_i}(x_i;e^*)dx_i}$
$e^*$ can be calculated by equation \eqref{roundk0} as a function of $x$.
\begin{equation}
\begin{split}
0={\partial L_2\over \partial e^*}=n(\int{f_{e_i}(x_i,e^*)x_idx_i})-nc'(e^*)\\
-q(c''(e^*)-k_0{\delta\over{1-\delta}}{\partial {\int_{t(e^*)}^\infty F(x_i,e^*)^{n-1}f_{e_i}(x_i;e^*)dx_i}\over \partial e^*})\\
=n(\int{f_{e_i}(x_i,e^*)x_idx_i}-c'(e^*))\\
-{\over\int_{t(e^*)}^\infty F(x_i,e^*)^{n-1}f_{e_i}(x_i;e^*)dx_i}({{1-\delta}\over\delta}c''(e^*)-k_0{\partial {\int_{t(e^*)}^\infty F(x_i,e^*)^{n-1}f_{e_i}(x_i;e^*)dx_i}\over \partial e^*})\\
\end{split}
\end{equation}
Therefore, by \eqref{t},
\begin{equation}
\begin{split}
{{(1-\delta)}\over\delta}{\partial\over\partial e^*}{c'(e^*)\over\int_{t(e^*)}^\infty F(x_i,e^*)^{n-1}f_{e_i}(x_i;e^*)dx_i}\\
={{(1-\delta)}\over\delta}({c''(e^*)\over\int_{t(e^*)}^\infty F(x_i,e^*)^{n-1}f_{e_i}(x_i;e^*)dx_i}\\
-{c'(e^*)\over(\int_{t(e^*)}^\infty F(x_i,e^*)^{n-1}f_{e_i}(x_i;e^*)dx_i)^2}{\partial {\int_{t(e^*)}^\infty F(x_i,e^*)^{n-1}f_{e_i}(x_i;e^*)dx_i}\over \partial e^*})\\
=n(\int{f_{e_i}(x_i,e^*)x_idx_i}-c'(e^*))
\end{split}
\end{equation}
In this case, for the same reason as with the inner contract, $c'(e^*)\leq{\delta\over{1-\delta}} (E(y;e^*)-nc(e^*)-n\bar{u}-\bar{u_0}[n]-k_0)\int_{\tilde{t}(e^*)}^\infty{g_{e_i}(y,e^*)dy}$.
\end{proof}

\subsection{Bonus and profit analysis} 
Then, I can analyze the profit for each party. First, by Lemma \ref{simp}, the profit of the workers is $0$. This is because the manager can reduce the salary until the profit becomes $0$. Note that the profit of the manager and the owner can be described as $k_0=\alpha_0+E(b_0)-\sum{c(e_i)}-\bar{u_0}[n]-n\bar{u}$ and $E(y;e^*)-nc(e^*)-n\bar{u}-\bar{u_0}[n]-k_0$, respectively.

By Theorem \ref{outsol}, the profit of the manager and the owner can be calculated as follows.
\begin{corollary}\label{profits}
The profit of the manager is decided as
\begin{equation}\label{managerprofit}
k_0={(1-\delta)c'(e^*)\over\delta\int_{t(e^*)}^\infty F(x_i,e^*)^{n-1}f_{e_i}(x_i;e^*)dx_i}.
\end{equation}
Additionally, if the sub-constraint of the outer contract holds, the profit of the owner can be determined as
\begin{equation}\label{ownerprofit}
{(1-\delta)c'(e^*)\over\delta\int_{\tilde{t}(e^*)}^\infty{g_{e_i}(y,e^*)dy}}.
\end{equation}
\end{corollary}

The proportion of surplus distribution can be calculated by Corollary \ref{profits} as follows.
\begin{corollary}\label{profitratio}
When the sub-constraint of the outer contract binds, the owner and the manager divide the total surplus according to the proportion
\begin{equation}
\int_{\tilde{t}(e^*)}^\infty{g_{e_i}(y,e^*)dy}:\int_{t(e^*)}^\infty F(x_i,e^*)^{n-1}f_{e_i}(x_i;e^*)dx_i.
\end{equation}
\end{corollary}
Note that the proportion in Corollary \ref{profitratio} is a function of only $n$, $f$, and $e^*$. This distribution is the inverse of the ratio of the ability of effort induced by giving a bonus.

Theorem \ref{differenceftn} presents the sufficient condition where the integrals in Corollary \ref{profitratio} can be independent of $e^*$.
\begin{theorem}\label{differenceftn}
For any constant $\xi$, when $f(x_i;e_i)$ depends only on $x_i-\xi e_i$, then $\int_{t(e^*)}^\infty F(x_i,e^*)^{n-1}f_{e_i}(x_i;e^*)dx_i=\int_{t(0)}^\infty F(x_i,0)^{n-1}f_{e_i}(x_i;0)dx_i$ and $\int_{\tilde{t}(e^*)}^\infty{g_{e_i}(y,e^*)dy}=\int_{\tilde{t}(0)}^\infty{g_{e_i}(y,0)dy}$.
\end{theorem}
\begin{proof}
By $f_{e_i}(t(e_i)-\xi e_i,0)=f_{e_i}(t(e_i),e_i)$, $t(0)=t(e^*)-\xi e^*$. Moreover, $F(x_i,0)=\int_{-\infty}^{x_i}f(a;0)da=\int_{-\infty}^{x_i}f(a+\xi e_i;e_i)da=\int_{-\infty}^{x_i+\xi e_i}f(\bar{a};0)d\bar{a}=F(x_i+\xi e_i,e_i)$.
Therefore, since $f_{e_i}(x_i;e_i)$ still depends on only $x_i-\xi e_i$,
\begin{equation}
\begin{split}
\int_{t(0)}^\infty F(x_i,0)^{n-1}f_{e_i}(x_i;0)dx_i=\int_{t(0)+\xi e^*}^\infty F(\tilde{x_i}-\xi e^*,0)^{n-1}f_{e_i}(\tilde{x_i}-\xi e^*;0)d\tilde{x_i}\\
=\int_{t(e^*)}^\infty F(\tilde{x_i},e^*)^{n-1}f_{e_i}(\tilde{x_i};e^*)d\tilde{x_i}.
\end{split}
\end{equation}
Furthermore,
\begin{equation}
g(y-n\xi e^*,0)=\int_{\sum_i(x_i+\xi e^*)=y}f(x_i,0)=\int_{\sum_i{\bar{x_i}}=y}f(\bar{x_i}-\xi e^*,0)=\int_{\sum_i{\bar{x_i}}=y}f(\bar{x_i},e^*)=g(y,e^*).
\end{equation}
Then, $\tilde{t}(0)=\tilde{t}(e^*)-n\xi e^*$ similar to the former logic, which leads to
\begin{equation}
\int_{\tilde{t}(0)}^\infty{g_{e_i}(y,0)dy}=\int_{\tilde{t}(e^*)-n\xi e^*}^\infty{g_{e_i}(y,0)dy}=\int_{\tilde{t}(e^*)}^\infty{g_{e_i}(\tilde{y}+n\xi e^*,0)d\tilde{y}}=\int_{\tilde{t}(e^*)}^\infty{g_{e_i}(\tilde{y},e^*)d\tilde{y}}.
\end{equation}
\end{proof}
Then, Corollary \ref{difference_ratio} is the direct result of Corollary \ref{profitratio} and Theorem \ref{differenceftn}.
\begin{corollary}\label{difference_ratio}
When the sub-constraint of the outer contract binds, the owner and the manager divide the total surplus according to the proportion
\begin{equation}
\int_{\tilde{t}(0)}^\infty{g_{e_i}(y,0)dy}:\int_{t(0)}^\infty F(x_i,0)^{n-1}f_{e_i}(x_i;0)dx_i,
\end{equation}
which only depends on $n$ and $f$.
\end{corollary}
Considering the concept of complementary slackness, the bonus scheme is forced to be optimal when the first-best result cannot be achieved. Thus, I can conclude that \textbf{under the optimal bonus scheme, the surplus is distributed to the owner and the manager at a constant rate $\int_{\tilde{t}(0)}^\infty{g_{e_i}(y,0)dy}:\int_{t(0)}^\infty F(x_i,0)^{n-1}f_{e_i}(x_i;0)dx_i$}.

\subsection{Simple solution under normal distribution}
I analyze the case where $f(x_i;e_i)$ is a normal distribution with mean $e_i$ and variance $\sigma^2$. Then, I can obtain Theorem \ref{p+rho}. This theorem proposes that the complex equation $\int_{\kappa}^\infty F(x_i,e^*)^{n-1}f_{e_i}(x_i;e^*)dx_i$ can be reduced to a function of only $n$ and $e^*-\kappa\over\sigma$ for all $\kappa$ regardless of $e^*$ when $f$ is a normal distribution. Note that the complementary error function $erfc(x)$ is defined as ${2\over\sqrt{\pi}}\int_x^\infty{e^{-t^2}dt}$.
\begin{theorem}\label{p+rho}
When $f$ is a normal distribution centered on $e^*$, $\forall \kappa$,
\begin{equation}
\int_{\kappa}^\infty F(x_i,e^*)^{n-1}f_{e_i}(x_i;e^*)dx_i={p_n+\rho_n(\eta)\over\sigma}.
\end{equation}
with $\eta={e^*-\kappa\over\sigma}$.
\end{theorem}
\begin{proof}
Given that
\begin{equation}
f(x,e)={1\over\sigma\sqrt{2\pi}}exp(-{(x-e)^2\over2\sigma^2}),
\end{equation}
I can simplify $\int_{\kappa}^\infty F(x_i,e^*)^{n-1}f_{e_i}(x_i;e^*)dx_i$ as follows.
\begin{equation}
\begin{split}
\int_{\kappa}^\infty F(x_i,e^*)^{n-1}f_{e_i}(x_i;e^*)dx_i=\int_{\kappa}^\infty F(x_i,e^*)^{n-1}{1\over\sigma\sqrt{2\pi}}{(x_i-e^*)\over\sigma^2}exp(-{(x_i-e^*)^2\over2\sigma^2})dx_i\\
=\int_{\kappa}^\infty F(x_i,e^*)^{n-1}f(x_i,e^*){(x_i-e^*)\over\sigma^2}dx_i\\
=\int_{e^*}^\infty F(x_i,e^*)^{n-1}f(x_i,e^*){(x_i-e^*)\over\sigma^2}dx_i+\int_{\kappa}^{e^*} F(x_i,e^*)^{n-1}f(x_i,e^*){(x_i-e^*)\over\sigma^2}dx_i\\
=\lim_{q\to\infty}({(F(q;e^*))^n\over n}{(q-e^*)\over\sigma^2}-\int_{e^*}^q{F(x_i,e^*)^n{1\over n\sigma^2}dx_i})-{(F(\kappa;e^*))^n\over n}{(\kappa-e^*)\over\sigma^2}-\int_{\kappa}^{e^*}{F(x_i,e^*)^n{1\over n\sigma^2}dx_i}\\
={1\over\sigma}{1\over n2^n}\lim_{w\to\infty}((erfc(-{w\over\sqrt2}))^nw-\int_0^w{(erfc(-{\bar{w}\over\sqrt{2}}))^nd\bar{w}})+{\eta erfc({\eta\over\sqrt{2}})^n\over n2^n\sigma}-\int_{0}^{\eta}{erfc({\tilde{w}\over\sqrt{2}}){1\over n2^n\sigma}d\tilde{w}}\\
={p_n+\rho_n(\eta)\over\sigma}
\end{split}
\end{equation}
\end{proof}

Since it is clear that $t(e^*)=e^*$ when $f$ is a normal distribution centered on $e^*$, I can obtain Corollary \ref{const} directly from Theorem \ref{p+rho}.
\begin{corollary}\label{const}
When $f$ is a normal distribution centered on $e^*$,
$\int_{t(e^*)}^\infty F(x_i,e^*)^{n-1}f_{e_i}(x_i;e^*)dx_i={p_n\over\sigma}$.
\end{corollary}

Moreover, $g(y,e_1,\dots,e_n)=N(\sum_i e_i,n\sigma^2)$. The properties of the normal distribution mentioned in this subsection help us obtain a simple solution of \eqref{popt}. The proof of Proposition \ref{simplified_sol} is given as follows.
\begin{proof}\label{proof_simp_sol}
The inner contract part is straightforward from Theorem \ref{T1} and Corollary \ref{const}.

For the outer contract part, because the expectation of $x$ is $e^*$, $\int{f_{e_i}(x_i,e^*)x_idx_i}=1$.
Substituting it and the result of Corollary \ref{const}, \eqref{genopte} becomes
\begin{equation}
{{(1-\delta)}\over\delta}{\partial\over\partial e^*}{c'(e^*)\sigma\over p_n}=n(1-c'(e^*))
\end{equation}
Since $p_n$ and $\sigma$ is independent to $e^*$, the equation can be rewritten as
\begin{equation}
{{\sigma(1-\delta)}\over\delta}{c''(e^*)\over np_n}=(1-c'(e^*)).
\end{equation}
By using $E(y;e^*)=ne^*$and Corollary \ref{const}, I can simplify the sub-constraint to
\begin{equation}
c'(e)\leq{1\over\sqrt{2\pi n}}({\delta\over({1-\delta})\sigma} (ny-nc(e^*)-n\bar{u}-\bar{u_0}[n])-{c'(e^*)\over p_n}).
\end{equation}
It can be easily rearranged to
\begin{equation}
c'(e^*)(\sqrt{2\pi\over n}+{1\over n p_n})\leq{\delta\over({1-\delta})\sigma} (e^*-c(e^*)-\bar{u}-{\bar{u_0}[n]\over n}).
\end{equation}
\end{proof}

\section{Detailed derivation of the optimal contract: Separate bonus scheme for the manager case}
\subsection{Constraints}\label{sep_constraints}
Similar to the former case, I use backward induction to analyze the contract. The contract between the manager and the workers and the contract between the owner and the team (the manager and the workers) are referred to as the \textit{inner contract} and \textit{outer contract}, respectively. I first derive the optimal inner contract and use it as a constraint to obtain the optimal outer contract. I consider only one time step for \eqref{IC_sep} (the same as in the former case) and \eqref{IC_M_sep} as in the previous section. First, the manager and workers maximize their profits, and the workers select the effort level that will achieve the maximum profit in one period.
\begin{equation}\label{IC_sep}
e_i=\argmax_{e_i'} (\alpha_i+E(b_i)-c(e_i'))
\end{equation}
The manager selects the bonus function $b_i(\bold{x})$ and wages $\alpha_i$ that maximize his or her profits. Since the salary of the manager is decided by the owner, he or she only has to maximize his or her bonus. That is,
\begin{equation}\label{IC_M_sep}
(\alpha_1, \dots, \alpha_n, b_1, \dots, b_n)=\argmax_{\alpha_1', \dots, \alpha_n',b_1', \dots, b_n'} E(b_m)
\end{equation}

Similar to the previous case, for the relational contract to be sustainable, the owner should be able to obtain more potential profit if he or she gives the appropriate bonus to the manager.
\begin{equation}\label{EP_sep}
\forall y, -b_m(y)+{\delta\over{1-\delta}} (E(y-b_m)-\alpha_m-\alpha_t-b_t)\geq0
\end{equation}

Assuming \eqref{IC}, workers continue the contract when they earn a profit greater than $\bar{u}$ from the contract.
\begin{equation}\label{EA_sep}
\alpha_i+E(b_i)-c(e_i)\geq\bar{u}
\end{equation}

The manager should be able to obtain more potential profit if he or she gives the appropriate bonus to the workers. When other workers can observe the corruption, the relational contract will break even though they cannot prove the corruption in court.

Since the minimal possible bonus $\min_{i,\bold{x}} b_i(\bold{x})>0$, it can be treated as salary because the total salary and the total bonus are indifferently and simultaneously paid by the owner. Then, I can assume that the minimal possible bonus for workers is $0$, and the manager and the corrupt worker can take $\phi b_t$.
\begin{equation}\label{EM_sep}
{\delta\over{1-\delta}}(\alpha_m+E(b_m)-\bar{u_0}[n]+\alpha_i+E(b_i)-c(e_i)-\bar{u})\geq \phi b_t
\end{equation}

Finally, there are constraints on salary and bonus distribution.
\begin{equation}\label{salary_sep}
\sum_i\alpha_i\leq\alpha_t
\end{equation}
\begin{equation}\label{bonus_sep}
\forall\bold{x}, \sum_ib_i(\bold{x})\leq b_t
\end{equation}
\subsection{inner contract}
The optimization problem for the contract between the manager and the workers is presented below.
\begin{subequations}\label{inner_ori_sep}
\begin{align}
& \max_{\alpha_1', \dots, \alpha_n',b_1', \dots, b_n'} &&E(b_m)\label{inner_oriobj_sep}\\
&\text{subject to}&&\eqref{IC_sep}, \eqref{EM_sep}, \eqref{EA_sep}, \eqref{salary_sep}, \eqref{bonus_sep}\\
& &&\forall i,\forall \bold{x}, b_i(\bold{x})\geq0
\end{align}
\end{subequations}

To simplify problem \eqref{inner_ori_sep}, I employ Assumption \ref{soc}.

\begin{lemma}\label{simp_sep}
I can let $\alpha_i={\alpha_t\over n}$ without loss of optimality. Furthermore, under Assumption \ref{soc}, the optimization problem \eqref{inner_ori_sep} can be reduced to \eqref{optbi_sep}.
\begin{subequations}\label{optbi_sep}
\begin{align}
& \max_{(b_1, \dots, b_n)} &&E(b_m)\\
&\text{subject to}&&\forall i, {\partial E(b_i)\over\partial e_i}=c'(e_i)\label{mu_sep}\\
& &&\forall\bold{x}, \sum_ib_i(\bold{x})\leq b_t\label{lambda_sep}\\
& &&\forall i, E(b_i)-c(e_i)\geq\bar{u}-{\alpha_t\over n}+({b_t\phi(1-\delta)\over\delta}-\alpha_m-E(b_m)+\bar{u_0}[n])^+\label{theta_sep} \\
& &&\forall i,\forall \bold{x}, b_i(\bold{x})\geq0\label{tau_sep}
\end{align}
\end{subequations}
\end{lemma}
\begin{proof}
By Assumption \ref{soc}, \eqref{IC_sep} can be replaced by \eqref{mu_sep}. Furthermore, since the only upper bound of $\alpha_i$ is \eqref{salary_sep} and symmetry, I can let $\alpha_i={\alpha_t\over n}$ without loss of optimality because the objective does not include $\alpha_i$.
\end{proof}
I employ Assumption \ref{mlrp} as in the previous section. The simplified optimization problem \eqref{optbi_sep} can be solved in a manner similar to the former case. The solution is presented in Theorem \ref{inner_sep}.

\begin{theorem}\label{inner_sep}
Let
\begin{equation}\label{tbar}
\bar{t}(e)\equiv\min(t(e), F^{-1}((1-{nc(e)+n\bar{u}-\alpha_t\over b_t}-n({\phi(1-\delta)\over\delta}-{\alpha_m+E(b_m)-\bar{u_0}[n]\over b_t})^+)^{1\over n};e)).
\end{equation}
Then, $b_i(x_i)=b_t$ when $x_i>\bar{t}(e^*)$ and $x_i$ is maximal and $b_i(x_i)=0$ otherwise.
Moreover, $e^*$ satisfies
\begin{equation}\label{sol_inner_sep}
c'(e^*)=b_t\int_{\bar{t}(e^*)}^\infty F(x_i,e^*)^{n-1}f_{e_i}(x_i;e^*)dx_i.
\end{equation}
\end{theorem}
\begin{proof}
The Lagrangian of \eqref{optbi_sep} can be written as follows. Let $\mu_i$, $\lambda(\bold{x})$, $\theta_i$, and $\tau_i(\bold{x})$ be Lagrange multipliers of \eqref{mu_sep}, \eqref{lambda_sep}, \eqref{theta_sep} and \eqref{tau_sep}. $i^-$ denotes $1,\dots,i-1,i+1,\dots,n$.
\begin{equation}\label{lanbi}
\begin{split}
L_1=E(b_m)+\sum_i{\int{\tau_i(\bold{x})b_i(\bold{x})d\bold{x}}}+\int{\lambda(\bold{x})(b_t-\sum_i{b_i(\bold{x})})d\bold{x}}+\\
\sum_i{\mu_i(\int{b_i(\bold{x})f_{e_i}(x_i;e_i)f(\bold{x}_{i^{-}};\bold{e}_{i^{-}})d\bold{x}}-c'(e_i))}
+\\
\sum_i\theta_i(\int{b_i(\bold{x})f(\bold{x})d\bold{x}}+{\alpha_t\over n}-c(e_i)-\bar{u}-({b_t\phi(1-\delta)\over\delta}-\alpha_m-E(b_m)+\bar{u_0}[n])^+)
\end{split}
\end{equation}
\begin{equation}\label{lanbi}
\begin{split}
{\partial L_1\over \partial e_i}={\partial E(b_m)\over \partial e_i}+\sum_j{\mu_j{\partial ({\partial E(b_j)\over\partial e_j}-c'(e_j))\over \partial e_i}}-\theta_ic'(e_i)
\end{split}
\end{equation}
Due to symmetry and Assumption \ref{soc}, I can assume all $\mu_i$ and optimal $e_i$ are the same over $i$.
Let $\mu_1=\mu_2=\dots=\mu_n\equiv\mu, \theta_1=\theta_2=\dots=\theta_n\equiv\theta, e_1^*=e_2^*=\dots=e_n^*\equiv e^*$.
\begin{equation}\label{lanbi}
\begin{split}
{\partial L_1\over \partial b_i(\bold{x})}=\mu f_{e_i}(x_i;e_i)f(\bold{x}_{i^{-}};\bold{e}_{i^{-}})-\lambda(\bold{x})+\tau_i(\bold{x})+\theta f(\bold{x})=(\mu {f_{e_i}(x_i;e_i)\over f(x_i;e_i)}+\theta)f(\bold{x})-\lambda(\bold{x})+\tau_i(\bold{x})
\end{split}
\end{equation}

Then, $\lambda(\bold{x})=f(\bold{x})\max_i{(\mu {f_{e_i}(x_i;e_i)\over f(x_i;e_i)}+\theta)}$ if $\exists i, (\mu {f_{e_i}(x_i;e_i)\over f(x_i;e_i)}+\theta)>0$ and $0$ otherwise. Therefore, $b_i=b_t$ and $\tau_i=0$ for $i\in\argmax_i{\mu {f_{e_i}(x_i;e_i)\over f(x_i;e_i)}}$ and $\tau_i>0, b_i=0$ for other $i$ when $\exists i, (\mu {f_{e_i}(x_i;e_i)\over f(x_i;e_i)}+\theta)>0$ and $\forall i, \tau_i>0, b_i=0$ when $\forall i, (\mu {f_{e_i}(x_i;e_i)\over f(x_i;e_i)}+\theta)<0$.

By Assumption \ref{mlrp}, I can find a unique $t(e^*,h)\in\mathbb{R}^\infty$ that satisfies ${f_{e_i}(x_i;e_i)\over f(x_i;e_i)}=h$. (I define $t(e^*,h)=\infty$ or $t(e^*,h)=-\infty$ if $\forall x_i, {f_{e_i}(x_i;e_i)\over f(x_i;e_i)}>h$ or $\forall x_i, {f_{e_i}(x_i;e_i)\over f(x_i;e_i)}<h$.) Note that $t(e^*,0)=t(e^*)$. If $\mu<0$, $\tau_i(\bold{x})>0$ and $b_i(x_i)=0$ when $x_i>t(e^*,-{\theta\over\mu})$. $\lambda(\bold{x})=f(\bold{x})\max_i{(\mu {f_{e_i}(x_i;e_i)\over f(x_i;e_i)}+\theta)}$ if $\exists i, x_i<t(e^*, -{\theta\over\mu})$. $b_i(x_i)=b_t\geq0$ when $x_i<t(e^*,-{\theta\over\mu})$ and $x_i$ is minimal. Clearly, this case leads to ${\partial E(b_i)\over\partial e_i}<0$ and collides with Assumption \ref{soc}.

The last case I consider is $\mu>0$. $\tau_i(\bold{x})>0$ and $b_i(x_i)=0$ when $x_i<t(e^*)$.
$\lambda(\bold{x})=f(\bold{x})\max_i{(\mu {f_{e_i}(x_i;e_i)\over f(x_i;e_i)}+\theta)}$ if $\exists i, x_i>t(e^*)$. $b_i(x_i)=b_t\geq0$ when $x_i>t(e^*)$ and $x_i$ is maximal.

Finally, I can use the complementary slackness of $\theta$. By $\theta(E(b_i)-c(e_i)-\bar{u}+{\alpha_t\over n})-({b_t\phi(1-\delta)\over\delta}-\alpha_m-E(b_m)+\bar{u_0}[n])^+=0$, I can conclude that ${b_t\over n}(1-F(t(e^*,-{\theta\over\mu}),e^*)^n)=b_t\int_{t(e^*,-{\theta\over\mu})}^\infty F(x_i,e^*)^{n-1}f(x_i;e^*)dx_i=E(b_i)=c(e^*)+\bar{u}-{\alpha_t\over n}+({b_t\phi(1-\delta)\over\delta}-\alpha_m-E(b_m)+\bar{u_0}[n])^+$ when $\theta>0$. Therefore, I can obtain $t(e^*,-{\theta\over\mu})=\min(t(e^*), F^{-1}((1-{nc(e)+n\bar{u}-\alpha_t\over b_t}-n({\phi(1-\delta)\over\delta}-{\alpha_m+E(b_m)-\bar{u_0}[n]\over b_t})^+)^{1\over n};e^*))$.

As a result, by \eqref{mu_sep}, $e^*$ satisfies $c'(e^*)=b_t\int_{\bar{t}(e^*)}^\infty F(x_i,e^*)^{n-1}f_{e_i}(x_i;e^*)dx_i$.
\end{proof}

In the previous case, the participation constraint of workers \eqref{EA} is merged with \eqref{EM} and has meaning only as a part of surplus because the manager can adjust the salary freely by changing his or her salary. Conversely, in this case, the manager has a strict restriction on distributing the salary and bonus. Theorem \ref{inner_sep} shows that the effect of participation constraint \eqref{theta_sep} is only to lower the threshold to give a bonus until the constraint holds instead of limiting the maximal bonus as in the previous case. This is a result of Assumption \ref{mlrp}.

\subsection{outer contract}
Then, I analyze the outer contract (the contract between the owner and the team) using the result of the inner contract as a constraint. The optimization problem to obtain the optimal outer contract can be written as follows.
\begin{subequations}\label{outer_prob_sep}
\begin{align}
& \max_{(\alpha_t, b_t, \alpha_m, b_m(y))} &&E(y-b_m)-\alpha_t-b_t-\alpha_m\\
&\text{subject to}&&c'(e^*)=b_t\int_{\bar{t}(e^*)}^\infty F(x_i,e^*)^{n-1}f_{e_i}(x_i;e^*)dx_i\label{inner_sol_sep}\\
& &&\forall y, b_m(y)\leq{\delta\over{1-\delta}} (E(y-b_m)-\alpha_m-\alpha_t-b_t)\label{bmmax_sep}\\
& &&\forall y, b_m(y)\geq0\label{bmmin_sep}\\
& &&\alpha_m+E(b_m)\geq\bar{u_0}[n]\label{feas_sep_outer}
\end{align}
\end{subequations}
I can solve the optimization problem \eqref{outer_prob_sep} as follows.
\begin{theorem}\label{outer_sep}
The optimal $e^*$ and $\bar{t}(e^*)$ associated with the optimal solution satisfy the two equations presented below.
\begin{equation}\label{estarfinal_sep}
\begin{split}
{f_{e_i}(\bar{t}(e^*);e^*)(\int{f_{e_i}(x_i,e^*)x_idx_i})\over f(\bar{t}(e^*),e^*)}
\\=c''(e^*)-{c'(e^*)\over\int_{\bar{t}(e^*)}^\infty F(x_i,e^*)^{n-1}f_{e_i}(x_i;e^*)dx_i}{\partial {\int_{\bar{t}(e^*)}^\infty F(x_i,e^*)^{n-1}f_{e_i}(x_i;e^*)dx_i}\over \partial e^*}|_{\alpha_t, b_t}
\end{split}
\end{equation}
\begin{equation}\label{btfinal_sep}
{f_{e_i}(\bar{t}(e^*);e^*)\over f(\bar{t}(e^*),e^*)}({\phi(1-\delta)\over\delta}+F(\bar{t}(e^*),e^*)^n)=-n\int_{\bar{t}(e^*)}^\infty {F(x_i,e^*)^{n-1}f_{e_i}(x_i;e^*)dx_i}
\end{equation}
When there are no solutions for \eqref{estarfinal_sep} and \eqref{btfinal_sep}, the optimal $\bar{t}(e^*)=-\infty$ and the optimal $e^*$ is the solution of
\begin{equation}\label{inftysol}
n-nc'(e^*)-{\phi(1-\delta)\over\delta\int_{-\infty}^\infty {F(x_i,e^*)^{n-1}f_{e_i}(x_i;e^*)dx_i}}c''(e^*)=0.
\end{equation}

Furthermore, $b_t$ and $\alpha_t$ can be obtained from
\begin{equation}\label{optbt_sep}
b_t={c'(e^*)\over\int_{\bar{t}(e^*)}^\infty F(x_i,e^*)^{n-1}f_{e_i}(x_i;e^*)dx_i},
\end{equation}
\begin{equation}\label{optalphat_sep}
\alpha_t=nc(e^*)+n\bar{u}-b_t(1-F(\bar{t}(e^*),e^*)^n).
\end{equation}
Finally, without of loss optimality, $b_m$ and $\alpha_m$ can be set as $\forall y, b_m(y)=0$ and $\alpha_m=\bar{u_0}[n]+{\phi(1-\delta)\over\delta}b_t$.
\end{theorem}
\begin{proof}
At first, since the manager does not need to exert his or her effort with cost, the owner can let $b_m\equiv0$ satisfy \eqref{bmmin_sep}, \eqref{feas_sep_outer} by giving minimal money to the manager. To be accurate, the owner only needs to give an infinitesimal bonus for high total performance and $0$ for low total performance. Then, \eqref{bmmax_sep} is satisfied when the owner can achieve positive profit. As a result, I only need to consider \eqref{inner_sol_sep} as a constraint that affects the optimal $\alpha_t$, $\alpha_m$, and $b_t$.

Note that both $\alpha_t$ and $\alpha_m$ affect the optimal effort level and the outcome only through \eqref{tbar}, in which $E(b_m)=0$ since $b_m\equiv0$. Although $\alpha_t$ and $\alpha_m$ are indifferent for the owner, $\alpha_m$ worth $n$ times more than $\alpha_t$ when ${\phi(1-\delta)\over\delta}>{\alpha_m-\bar{u_0}[n]\over b_t}$ and the marginal increase of $\alpha_m$ is meaningless for the team when ${\phi(1-\delta)\over\delta}<{\alpha_m-\bar{u_0}[n]\over b_t}$. Therefore, I can conclude that the optimal $\alpha_m$ is given as $\alpha_m={\phi(1-\delta)b_t\over\delta}+\bar{u_0}[n]$.

Let $\zeta$ denote the Lagrange multiplier associated with \eqref{inner_sol_sep}. Then the Lagrangian can be written as follows.
\begin{equation}
L_2=E(y)-\alpha_t-{\delta+\phi(1-\delta)\over\delta}b_t-\bar{u_0}[n]+\zeta(c'(e^*)-b_t\int_{\bar{t}(e^*)}^\infty {F(x_i,e^*)^{n-1}f_{e_i}(x_i;e^*)dx_i})
\end{equation}
I can obtain equations about $\alpha_t, b_t, \zeta$ by differentiating the Lagrangian.
\begin{equation}
0={\partial L_2\over \partial e^*}=n(\int{f_{e_i}(x_i,e^*)x_idx_i})
+\zeta(c''(e^*)-b_t{\partial {\int_{\bar{t}(e^*)}^\infty F(x_i,e^*)^{n-1}f_{e_i}(x_i;e^*)dx_i}\over \partial e^*})
\end{equation}
\begin{equation}\label{diffalpha_sep}
0={\partial L_2\over \partial \alpha_t}=-1
-\zeta b_t{\partial \bar{t}(e^*)\over \partial \alpha_t}{\partial {\int_{\bar{t}(e^*)}^\infty F(x_i,e^*)^{n-1}f_{e_i}(x_i;e^*)dx_i}\over \partial \bar{t}(e^*)}=\zeta b_t{\partial \bar{t}(e^*)\over \partial \alpha_t}F(\bar{t}(e^*),e^*)^{n-1}f_{e_i}(\bar{t}(e^*);e^*)-1
\end{equation}
\begin{equation}
0={\partial L_2\over \partial b_t}=\zeta b_t{\partial \bar{t}(e^*)\over \partial b_t}F(\bar{t}(e^*),e^*)^{n-1}f_{e_i}(\bar{t}(e^*);e^*)-\zeta\int_{\bar{t}(e^*)}^\infty {F(x_i,e^*)^{n-1}f_{e_i}(x_i;e^*)dx_i}-{\delta+\phi(1-\delta)\over\delta}
\end{equation}
By Assumption \ref{mlrp}, $|f_{e_i}(\bar{t}(e^*);e^*)|$ is sufficiently small when $\bar{t}(e^*)-t(e^*)$ is sufficiently small. Therefore, by \eqref{diffalpha_sep}, I only need to consider the case where $\bar{t}(e^*)=F^{-1}((1-{nc(e)+n\bar{u}-\alpha_t\over b_t})^{1\over n};e^*)$.
For simplicity, I use $e^*,\bar{t}(e^*)$ as the main variables. Then, $b_t={c'(e^*)\over\int_{\bar{t}(e^*)}^\infty F(x_i,e^*)^{n-1}f_{e_i}(x_i;e^*)dx_i}$, ${\partial \bar{t}(e^*)\over \partial \alpha_t}={1\over nb_tF(\bar{t}(e^*),e^*)^{n-1}f(\bar{t}(e^*),e^*)}$, and ${\partial \bar{t}(e^*)\over \partial b_t}={1-F(\bar{t}(e^*),e^*)^n\over nb_tF(\bar{t}(e^*),e^*)^{n-1}f(\bar{t}(e^*),e^*)}$.

Substituting them, I can obtain
\begin{equation}\label{estardiff_sep}
n(\int{f_{e_i}(x_i,e^*)x_idx_i})
+\zeta(c''(e^*)-{c'(e^*)\over\int_{\bar{t}(e^*)}^\infty F(x_i,e^*)^{n-1}f_{e_i}(x_i;e^*)dx_i}{\partial {\int_{\bar{t}(e^*)}^\infty F(x_i,e^*)^{n-1}f_{e_i}(x_i;e^*)dx_i}\over \partial e^*})=0,
\end{equation}
\begin{equation}\label{zetasub}
{\zeta f_{e_i}(\bar{t}(e^*);e^*)\over nf(\bar{t}(e^*),e^*)}=-1,
\end{equation}
\begin{equation}\label{btdiff_sep}
{\zeta f_{e_i}(\bar{t}(e^*);e^*)(1-F(\bar{t}(e^*),e^*)^n)\over nf(\bar{t}(e^*),e^*)}=\zeta\int_{\bar{t}(e^*)}^\infty {F(x_i,e^*)^{n-1}f_{e_i}(x_i;e^*)dx_i}+{\delta+\phi(1-\delta)\over\delta}.
\end{equation}
By substituting \eqref{zetasub} into \eqref{estardiff_sep} and \eqref{btdiff_sep}, I can remove $\zeta$.
\begin{equation}
\begin{split}
{f_{e_i}(\bar{t}(e^*);e^*)(\int{f_{e_i}(x_i,e^*)x_idx_i})\over f(\bar{t}(e^*),e^*)}\\
=c''(e^*)-{c'(e^*)\over\int_{\bar{t}(e^*)}^\infty F(x_i,e^*)^{n-1}f_{e_i}(x_i;e^*)dx_i}{\partial {\int_{\bar{t}(e^*)}^\infty F(x_i,e^*)^{n-1}f_{e_i}(x_i;e^*)dx_i}\over \partial e^*}|_{\alpha_t,b_t}
\end{split}
\end{equation}
\begin{equation}
{f_{e_i}(\bar{t}(e^*);e^*)\over f(\bar{t}(e^*),e^*)}({\phi(1-\delta)\over\delta}+F(\bar{t}(e^*),e^*)^n)=-n\int_{\bar{t}(e^*)}^\infty {F(x_i,e^*)^{n-1}f_{e_i}(x_i;e^*)dx_i}
\end{equation}

When these equations do not have a solution, the optimal $\bar{t}(e^*)$ is not a real number. Since ${\alpha_t\over n}+E(b_i)\geq c(e)+\bar{u}$ to the contract feasible, then optimal $\bar{t}(e^*)$ is $-\infty$ and $\alpha_t+b_t=nc(e)+n\bar{u}$. Substituting them into the original problem \eqref{outer_prob_sep}, I obtain new objective
\begin{equation}\label{newobj_out_sep}
E(y)-{\phi(1-\delta)\over\delta}b_t-n\bar{u}-nc(e^*)-\bar{u_0}[n]
\end{equation}
and new constraint
\begin{equation}
c'(e^*)=b_t\int_{-\infty}^\infty {F(x_i,e^*)^{n-1}f_{e_i}(x_i;e^*)dx_i}.
\end{equation}
Using the new Lagrange multiplier $\bar{\zeta}$, I can write a new Lagrangian
\begin{equation}\label{newobj_out_sep}
\bar{L_2}=E(y)-{\phi(1-\delta)\over\delta}b_t-n\bar{u}-nc(e^*)-\bar{u_0}[n]+\bar{\zeta}(c'(e^*)-b_t\int_{-\infty}^\infty {F(x_i,e^*)^{n-1}f_{e_i}(x_i;e^*)dx_i})
\end{equation}
Differentiating it with $e^*$ and $b_t$, I can obtain $n(\int{f_{e_i}(x_i,e^*)x_idx_i})-nc'(e^*)+\bar{\zeta}c''(e^*)=0$, ${\phi(1-\delta)\over\delta}+\bar{\zeta}\int_{-\infty}^\infty {F(x_i,e^*)^{n-1}f_{e_i}(x_i;e^*)dx_i}=0$.

Therefore, $e^*$ is the solution of
\begin{equation}
n(\int{f_{e_i}(x_i,e^*)x_idx_i})-nc'(e^*)-{\phi(1-\delta)\over\delta\int_{-\infty}^\infty {F(x_i,e^*)^{n-1}f_{e_i}(x_i;e^*)dx_i}}c''(e^*)=0.
\end{equation}
\end{proof}
Note that $e^*$ and $\bar{t}(e^*)$ are the solutions of the system of equations consisting of \eqref{estarfinal_sep} and \eqref{btfinal_sep}. I cannot further solve the system of equations because of its complexity. This outer contract has \eqref{estarfinal_sep} and \eqref{btfinal_sep} as the optimal equations. Although the outer contract seems to have two optimal equations, since these equations compose a system of equations for $e^*$ and $e^*-\sigma\eta$, the effective number of the optimal equation of the outer contract is one.

\subsection{Bonus and profit analysis}\label{bonus_profit_sep}
With the final part of Theorem \ref{inner_sep} and \eqref{optalphat_sep}, the profit of the workers $E(b_i)+c(e_i)-\bar{u}+{\alpha_t\over n}$ is always $0$. Additionally, by the time the final part of Theorem \ref{outer_sep} is reached, the profit of the manager is
\begin{equation}
{\phi(1-\delta)b_t\over\delta}={\phi(1-\delta)c(e^*)\over\delta\int_{\bar{t}(e^*)}^\infty F(x_i,e^*)^{n-1}f_{e_i}(x_i;e^*)dx_i}.
\end{equation}
The expected profit of the owner is $E(y-b_m(y))-b_t-\alpha_t-\alpha_m$. By \eqref{optbt_sep} and \eqref{optalphat_sep}, it can be expressed as
\begin{equation}
\begin{split}
\int_{-\infty}^{\infty}{yg(y;ne^*)dy}-nc(e^*)-b_tF(\bar{t}(e^*),e^*)^n-n\bar{u}-\bar{u_0}[n]\\
=\int_{-\infty}^{\infty}{yg(y;ne^*)dy}-nc(e^*)-{(\delta F(\bar{t}(e^*),e^*)^n+\phi(1-\delta))c'(e^*)\over\delta\int_{\bar{t}(e^*)}^\infty F(x_i,e^*)^{n-1}f_{e_i}(x_i;e^*)dx_i}-n\bar{u}-\bar{u_0}[n].
\end{split}
\end{equation}
The first term and the second term are the performance and the cost, respectively. The last two terms are the outside options. The third (middle) term indicates surplus destruction and the manager's profit. The total team bonus $b_t$ is destroyed when all workers perform under $\bar{t}(e^*)$. Then, the probability of surplus destruction is $F(\bar{t}(e^*),e^*)^n$.

Theorem \ref{differenceftn_sep} presents the sufficient condition where the bonus-winning probability can be independent of $e^*$.
\begin{theorem}\label{differenceftn_sep}
For any constant $\xi$, when $f(x_i;e_i)$ depends only on $x_i-\xi e_i$, then $\xi e^*-\bar{t}(e^*)=-\bar{t}(0)$ and $\int_{\bar{t}(e^*)}^\infty F(x_i,e^*)^{n-1}f_{e_i}(x_i;e^*)dx_i=\int_{\bar{t}(0)}^\infty F(x_i,0)^{n-1}f_{e_i}(x_i;0)dx_i$.
\end{theorem}
\begin{proof}
By $f_{e_i}(t(e_i)-\xi e_i,0)=f_{e_i}(t(e_i),e_i)$ and $F(x_i,0)=F(x_i+\xi e_i,e_i)$,
\begin{equation}
\begin{split}
\int_{\bar{t}(e^*)-\xi e^*}^\infty F(x_i,0)^{n-1}f_{e_i}(x_i;0)dx_i=\int_{\bar{t}(e^*)}^\infty F(\tilde{x_i}-\xi e^*,0)^{n-1}f_{e_i}(\tilde{x_i}-\xi e^*;0)d\tilde{x_i}\\
=\int_{\bar{t}(e^*)}^\infty F(\tilde{x_i},e^*)^{n-1}f_{e_i}(\tilde{x_i};e^*)d\tilde{x_i}.
\end{split}
\end{equation}
Then, \eqref{btfinal_sep} becomes
\begin{equation}
{f_{e_i}(\bar{t}(e^*)-\xi e^*;0)\over f(\bar{t}(e^*)-\xi e^*,0)}({\phi(1-\delta)\over\delta}F(\bar{t}(e^*)-\xi e^*,0)^n)=-n\int_{\bar{t}(e^*)-\xi e^*}^\infty {F(x_i,0)^{n-1}f_{e_i}(x_i;0)dx_i}
\end{equation}
Thus, $\bar{t}(e^*)-\xi e^*$ can be decided as the solution of the above equation, which only depends on ${\phi(1-\delta)\over\delta}$, $n$, and $f$, which means $\xi e^*-\bar{t}(e^*)=-\bar{t}(0)$.
Moreover, I can obtain
\begin{equation}
\int_{\bar{t}(e^*)}^\infty F(\tilde{x_i},e^*)^{n-1}f_{e_i}(\tilde{x_i};e^*)d\tilde{x_i}=\int_{\bar{t}(0)}^\infty {F(x_i,0)^{n-1}f_{e_i}(x_i;0)dx_i}.
\end{equation}
\end{proof}
By Theorem \ref{differenceftn_sep}, when the performance follows a fixed distribution around the constant multiplication of the effort, the difference between the constant multiplication of the effort and the threshold of giving the bonus is the function of only ${\phi(1-\delta)\over\delta}$, $f$, and $n$. Then, the bonus-winning probability and the probability of total team bonus destruction are also given as the function of only ${\phi(1-\delta)\over\delta}$, $f$, and $n$.

\subsection{Simple solution under normal distribution}
The proof of Proposition \ref{normalouter_sep} is given as follows.
\begin{proof}
In this proof, I use $\eta$ as a variable instead of $\bar{t}(e^*)$.
Using $f(x,e)={1\over\sigma\sqrt{2\pi}}exp(-{(x-e)^2\over2\sigma^2})$, I obtain the following equations.
\begin{equation}
{{f_{e_i}(\bar{t}(e^*);e^*)\over f(\bar{t}(e^*),e^*)}}={\bar{t}(e^*)-e^*\over\sigma^2}=-{\eta\over\sigma}, F(\bar{t}(e^*),e^*)={erfc({\eta\over\sqrt{2}})\over2}
\end{equation}
Moreover, by $E(x_i)=e_i$, it is clear that $\int{f_{e_i}(x_i,e^*)x_idx_i}=1$.

Then, by Theorem \ref{p+rho},
\begin{equation}
\int_{\bar{t}(e^*)}^\infty F(x_i,e^*)^{n-1}f_{e_i}(x_i;e^*)dx_i=\int_{e^*-\sigma\eta}^\infty F(x_i,e^*)^{n-1}f_{e_i}(x_i;e^*)dx_i={p_n+\rho_n(\eta)\over\sigma}
\end{equation}

Substituting them, the inner contract part is straightforward, and \eqref{estarfinal_sep}, and \eqref{btfinal_sep} can be simplified as
\begin{equation}
-{\eta\over\sigma}
=c''(e^*)-{c'(e^*)\sigma\over{p_n+\rho_n(\eta)}}{\partial {p_n+\rho_n(\eta)\over\sigma}\over \partial e^*}=c''(e^*)-{c'(e^*)\over{p_n+\rho_n(\eta)}}{\partial \eta\over \partial e^*}|_{\alpha_t, b_t}{\partial {\rho_n(\eta)}\over \partial \eta},
\end{equation}
\begin{equation}
-{\eta\over\sigma}({\phi(1-\delta)\over\delta}+{erfc({\eta\over\sqrt{2}})^n\over2^n})=-n{p_n+\rho_n(\eta)\over\sigma}.
\end{equation}
Substituting the definition of $\rho_n$, these equations become
\begin{equation}
-{\eta\over\sigma}
=c''(e^*)-{c'(e^*)\over{p_n+\rho_n(\eta)}}{\partial \eta\over \partial e^*}|_{\alpha_t, b_t}({1\over n2^n}(erfc({\eta\over\sqrt{2}})^n-n\eta ({1\over\sqrt{2\pi}})e^{-\eta^2\over2}erfc({\eta\over\sqrt{2}})^{n-1}-erfc({\eta\over\sqrt{2}})^n)),
\end{equation}
\begin{equation}
-{\eta\over\sigma}({\phi(1-\delta)\over\delta}+{erfc({\eta\over\sqrt{2}})^n\over2^n})=-n{p_n+{1\over n2^n}(\eta erfc({\eta\over\sqrt{2}})^n-\int_{0}^{\eta}{erfc({\tilde{w}\over\sqrt{2}})^nd\tilde{w}})\over\sigma}.
\end{equation}
These equations can be simplified as
\begin{equation}\label{opte_norm_sep}
{\eta\over\sigma}+c''(e^*)
=c'(e^*){\eta ({1\over\sqrt{2\pi}})e^{-\eta^2\over2}erfc({\eta\over\sqrt{2}})^{n-1}\over{2^n(p_n+\rho_n(\eta))}}{\partial \eta\over \partial e^*}|_{\alpha_t, b_t},
\end{equation}
\begin{equation}\label{opteta_proof}
2^n(np_n-\eta{\phi(1-\delta)\over\delta})=\int_{0}^{\eta}{erfc({\tilde{w}\over\sqrt{2}})^nd\tilde{w}}.
\end{equation}
When $\phi=0$, by the middle part of Theorem \ref{outer_sep}, the optimal $\eta$ is $\infty$, and the optimal effort satisfies \eqref{inftysol}. Thus, I obtain $c'(e^*)=1$, and the first best effort is achieved.

Otherwise, $\eta$ is given as the solution of \eqref{opteta_proof}.

By partially differentiating $b_t(1-{erfc({\eta\over\sqrt{2}})^n\over2^n})+\alpha_t=nc(e^*)+n\bar{u}$ by $\eta$, I obtain
\begin{equation}
{nb_te^{-{\eta^2\over2}}erfc({\eta\over\sqrt{2}})^{n-1}\over2^n\sqrt{2\pi}}=nc'(e^*){\partial e^*\over \partial \eta}|_{\alpha_t, b_t}.
\end{equation}
Therefore,
\begin{equation}
{\partial \eta\over \partial e^*}|_{\alpha_t, b_t}={2^n\sqrt{2\pi}c'(e^*)\over b_te^{-{\eta^2\over2}}erfc({\eta\over\sqrt{2}})^{n-1}}
={2^n\sqrt{2\pi}\int_{\bar{t}(e^*)}^\infty F(x_i,e^*)^{n-1}f_{e_i}(x_i;e^*)dx_i\over e^{-{\eta^2\over2}}erfc({\eta\over\sqrt{2}})^{n-1}}
={2^n\sqrt{2\pi}(p_n+\rho_n(\eta))\over \sigma e^{-{\eta^2\over2}}erfc({\eta\over\sqrt{2}})^{n-1}}
\end{equation}
Substituting this in \eqref{opte_norm_sep}, I obtain
\begin{equation}
{\eta\over\sigma}+c''(e^*)={c'(e^*)\eta\over\sigma}
\end{equation}
\end{proof}

\section{Proof of Proposition \ref{unobserve_choice}}
\begin{proof}
Based on the result of SubSection \ref{withoutmanager}, the first best $c'(e^*)=1$ is valid for a sufficiently small $\sigma$ in the equal bonus case. (When $\bar{u}>e^*-c(e^*)$ of the first best, all contracts are always infeasible.) Then, the owner profit becomes $e^*-c(e^*)-\bar{u}$, and this is the maximum profit that can be achieved in this situation. Thus, giving an equal bonus based on total performance without employing a manager is the best choice.

The subconstraint (of outer contract) of the case where employing a manager who distributes the salary and the bonus can be rewritten as follows:
\begin{equation}\label{subcon_inte_flatten}
ne^*-nc(e^*)-{\sigma(1-\delta)\over\delta}c'(e^*)(\sqrt{2\pi n}+{1\over p_n})-\bar{u_0}[n]-n\bar{u}\geq0
\end{equation}
It can be easily checked numerically that
\begin{equation}
\forall \delta, \phi, \sigma, \sqrt{2\pi n}+{1\over p_n}\geq {({\delta\over(1-\delta)}{erfc({\eta\over\sqrt{2}})^n\over2^n}+\phi) \over p_n+\rho_n(\eta)}.
\end{equation}
(Note that $\eta$, which is given as the solution of \eqref{opteta}, increases quickly when $\delta$ increases.)
Since the contract is feasible until the owner's profit \eqref{ownerprofit_sep} becomes $0$ in the case where employing a manager and giving a separate salary to the manager, it is clear that the subconstraint of the integrated bonus case is stricter than the feasibility condition of the separate bonus contract. Therefore, the separate bonus contract can be feasible until $\sigma$ is greater than the integrated bonus case.

To obtain the second statement, I compare the case where the owner gives an equal bonus with the case where the owner employs a manager and gives separate income to him or her. Since $n\sqrt{2\pi n}>{({\delta\over(1-\delta)}{erfc({\eta\over\sqrt{2}})^n\over2^n}+\phi) \over p_n+\rho_n(\eta)}$, by the same logic as in the former paragraph, the subconstraint of the integrated bonus case is stricter than the feasibility condition of the separate bonus contract when $\bar{u_0}[n]=0$. Let $\sigma_n^*$ denote the maximum feasible $\sigma$ for the equal bonus case. The separate income case with a manager can be feasible under greater $\sigma$ values than the equal bonus case without a manager if and only if a separate income case is feasible on $\sigma_n^*$. Let $U_n(\phi)$ denote
\begin{equation}
ne^*-nc(e^*)-(\sigma{erfc({\eta\over\sqrt{2}})^n\over2^n}+{\sigma(1-\delta)\over\delta}\phi) {c'(e^*)\over(p_n+\rho_n(\eta))}-n\bar{u}.
\end{equation}
($\eta$ and $e^*$ are the solutions to \eqref{opteta} and \eqref{opteqn_sep_normal}, respectively.)

Then, the separate income case is feasible on $\sigma_n^*$ for $\bar{u_0}\leq U_n(\phi)$. Moreover, since both $\eta$ and $e^*$ are decreasing $\phi$, $U_n(\phi)$ is a decreasing function of $\phi$.

Last, by former results, there exists a $\sigma=\tilde{\sigma}$ in which the owner profit of the equal bonus case and the separate bonus with manager case is the same. Since the optimal equation of the equal bonus case is the first best, when $\sigma=\tilde{\sigma}$, the subconstraint of the equal bonus case meets with the optimal equation of the separate bonus with the manager case. Comparing the subconstraints of two contracts including a manager, there exists $\tilde{\bar{u_0}}$ that causes the subconstraint of the integrated bonus case with a manager to outperform the subconstraint of the equal bonus case. Meanwhile, because the optimal equation of the integrated bonus case with a manager outperforms the separate bonus case when $\phi=1$ and the owner profit is continuous for $\phi$, when $\phi$ is sufficiently large, the optimal equation of the integrated bonus case outperforms the separate bonus case at $\sigma=\tilde{\sigma}$. Therefore, an integrated bonus contract with a manager is optimal when $\sigma=\tilde{\sigma}$ under $\bar{u_0}\leq\tilde{\bar{u_0}}$.
\end{proof}
\section{Numerical values about $p_n$ and $\rho_n(\eta)$}
The graphs of $p_n$, $p_n\sqrt{2\pi n}$, ans $np_n$ as a function of $n$ is presented below. 
\begin{figure}[h]
\centering
\includegraphics[width=\textwidth]{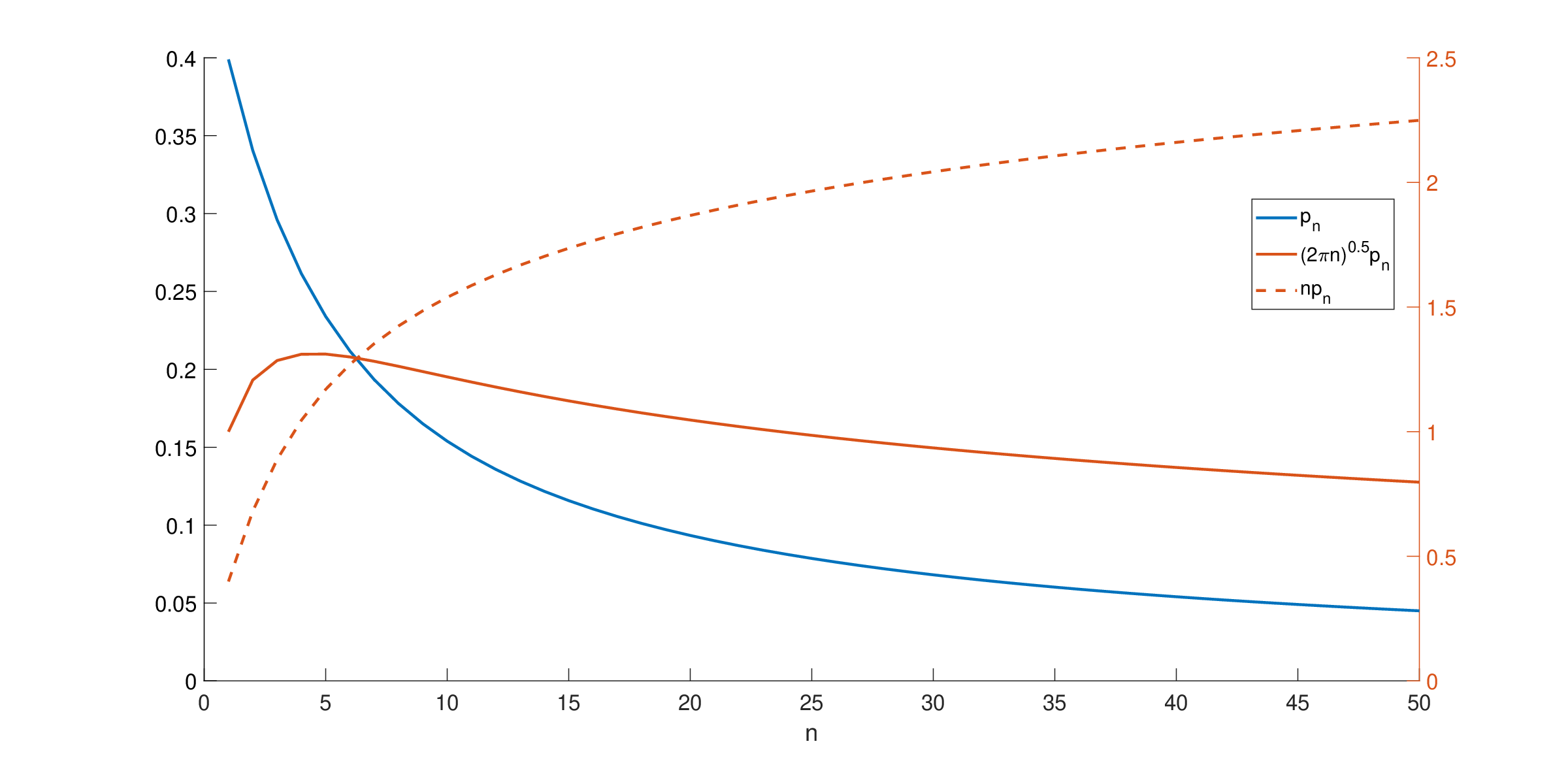}
\caption{$n$ vs $p_n$, $p_n\sqrt{2\pi n}$, $np_n$ graph}
\end{figure}

The graphs of $\rho_n(\eta)$ for some $n$ as a function of $\eta$ is presented below.
\begin{figure}[h]
\centering
\includegraphics[width=\textwidth]{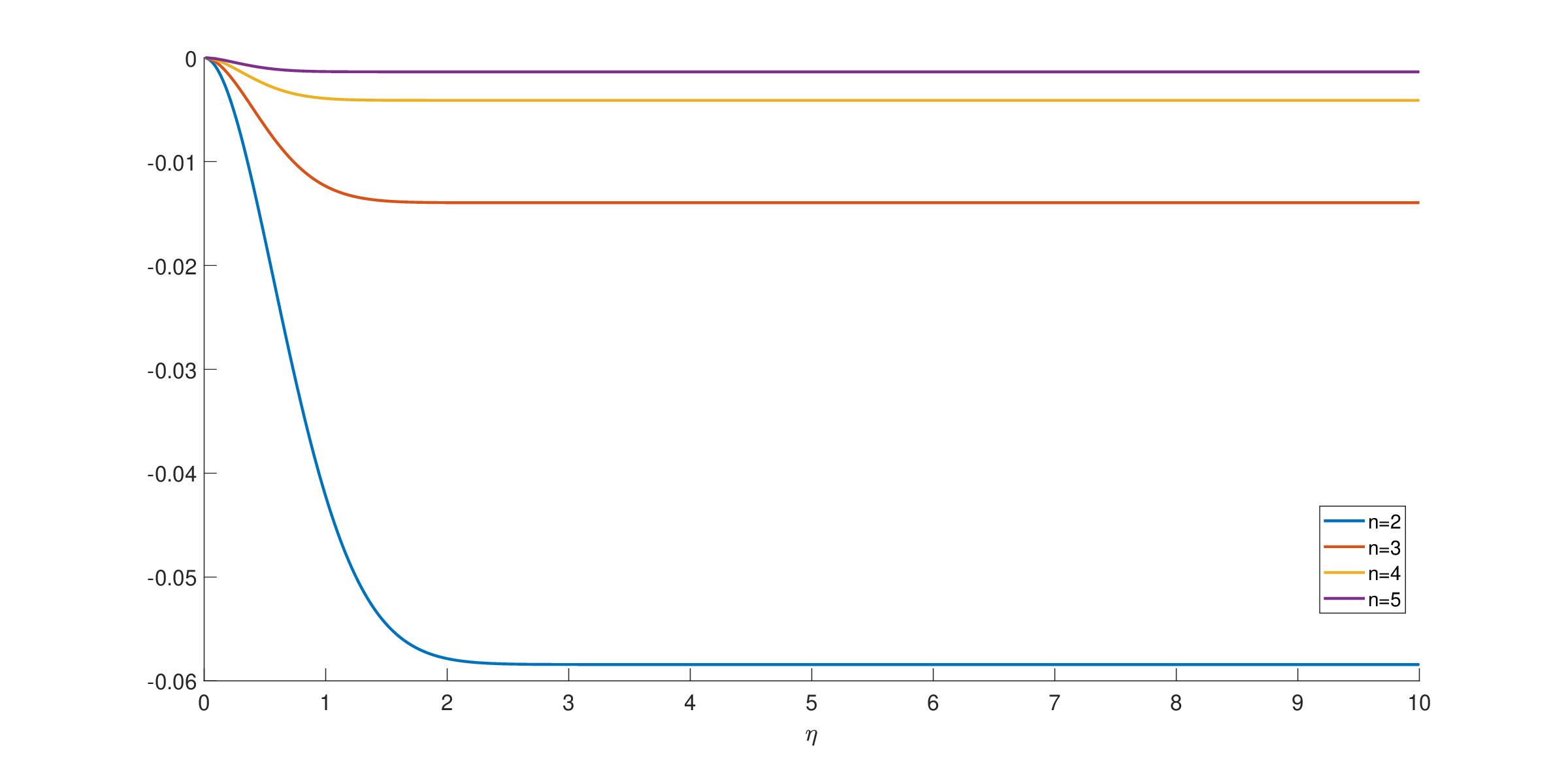}
\caption{$\rho_n(\eta)$ graph for $n=2,3,4,5$}
\end{figure}
\end{appendix}

\singlespacing
\bibliographystyle{chicago}

\bibliography{bib}

\end{document}